\documentclass[prb,twocolumn,showpacs,showkeys]{revtex4}
\usepackage{mathptmx}
\usepackage{amsfonts,float}
\usepackage{epsfig}
\newcommand{\beq}{\begin{equation}}
\newcommand{\eeq}{\end{equation}}
\newcommand{\beqa}{\begin{eqnarray}}
\newcommand{\eeqa}{\end{eqnarray}}
\newcommand{\nn}{\nonumber \\}
\def \e {\mathrm{e}}
\def \eps {\varepsilon}
\def \k {\kappa}
\def \l {\lambda}
\def \s {\sigma}
\def \t {\tau}
\def \C {{\mathbb C}}
\def \R {{\mathbb R}}
\def \Z {{\mathbb Z}}
\def \ch {\mathrm{ch}}
\def \qh {\mathrm{qh}}
\def \qp {\mathrm{qp}}
\def \z {\zeta}
\def \L {\underline{\Lambda}}
\def \D {\Delta}
\def \PF {\mathrm{PF}}
\def \Im {\mathrm{Im} \, }
\def \Re {\mathrm{Re} \, }%
\def \mod {\ \mathrm{mod} \ }
\def \H {{\mathcal H}}

\begin{document}
\title{Chiral persistent currents and magnetic susceptibilities in the 
parafermion quantum Hall states in the second Landau level with 
Aharonov--Bohm flux}
\author{\firstname{Lachezar} S. \surname{Georgiev}}
\email{lgeorg@inrne.bas.bg}
\affiliation{Institute for Nuclear Research and
Nuclear Energy, Tsarigradsko Chaussee 72,  1784 Sofia, BULGARIA\\ }

\begin{abstract}
Using the effective conformal field theory for the quantum Hall edge states we
propose a compact and convenient scheme for the computation of the
periods, amplitudes and temperature behavior  of the chiral persistent 
currents and the magnetic susceptibilities in the
mesoscopic disk version of the   $\Z_k$ parafermion  
quantum Hall states  in the second Landau level. 
Our numerical calculations show that the  persistent currents are
periodic in the Aharonov--Bohm flux with period exactly one flux quantum and 
have a  diamagnetic nature.
In the high-temperature regime their amplitudes decay exponentially 
with increasing the temperature  and the corresponding  exponents 
are  universal characteristics of non-Fermi liquids. 
Our theoretical results for these exponents  are in perfect 
agreement with those extracted  from the numerical data
and demonstrate that there is in general  a non-trivial contribution coming 
from the neutral  sector.
We emphasize the crucial role of the non-holomorphic
factors, first proposed by Cappelli and Zemba in the context of the 
conformal field theory partition functions  for the quantum Hall states, 
which ensure the invariance
of the annulus partition function under the Laughlin spectral 
flow. 
\end{abstract}
\maketitle
\section{Introduction}
According to a well-known but unpublished  theorem due to  Bloch
\cite{thouless:top},
the free energy of a conducting ring (or other non-simply connected
conductor) is a periodic function of the magnetic flux through the
ring with  period  one flux quantum $\phi_0=h/e=1$.
The flux dependence of the free energy
$F(T,\phi)=-k_B T \ln(Z(T,\phi))$, where $Z(T,\phi)$ is the partition
function, gives rise to  a non-dissipative  equilibrium current
\[
I=-\left(\frac{e}{h}\right) \frac{\partial F(T,\phi)}{\partial \phi}  , 
\]
flowing along the ring, called a \textit{persistent current},
which has a universal amplitude
and universal temperature dependence.
Such currents have been  experimentally observed in conducting  
mesoscopic rings \cite{pers-exp},
in which the circumference  of the ring is smaller than the  
coherence length  \cite{geller:encyclop}.
Mesoscopic effects, such as the persistent currents
and Aharonov--Bohm (AB) magnetization,  
offer an outstanding possibility for the experimental observation
of  certain implications of Quantum Mechanics.
This has recently become a topical issue in the context of
Quantum Computation \cite{QC}.
Another  motivation for our interest in mesoscopic physics is
that despite being equilibrium quantities
the persistent currents may give important information about transport 
due to a recently emerging relation between the Thouless energy, 
diagonal conductivity and the persistent currents\cite{altland,deo}. 
At the same time, these phenomena  could help us understand 
the effects of  electron--electron interaction on the strongly correlated 
electron systems  \cite{chakra-pietl-1,chakra-pietl-2}.

The fractional quantum Hall (FQH) effect 
provides an  exciting possibility for an experimental realization 
of mesoscopic rings\cite{geller:encyclop,geller-loss,geller-loss-kircz,ino}. 
Due to the non-zero energy gap in the bulk, when the system is on a FQH
 plateau,  the only delocalized states, which are able to carry electric 
current,  live on the edge of the FQH sample.
When this sample has the shape of a disk, its edge is a circular ring 
whose width is of the order of the magnetic length $l_B=\sqrt{\hbar/eB}$. 
For example, the edge states of the Laughlin FQH states gave  the first
observable  realizations of the chiral Luttinger 
liquids \cite{geller:encyclop}, whose mesoscopic properties have been
well-understood.

However, recent experiments \cite{grayson} have shown that
the Luttinger liquid theory is not relevant for more general filling 
factors such as those in the principle Jain series.
Fortunately, there is a more general and promising classification scheme 
for the FQH states based on the conformal field theory (CFT).
Within this scheme the universality classes of the FQH systems have been 
described by the effective field 
theories for the edge states in the (thermodynamic) scaling
limit \cite{fro-ker,fro-stu-thi,fro2000}. These turned out to be
Chern--Simons topological field theories in $2+1$ dimensional space-time 
which are equivalent to  unitary CFTs operating  on the $1+1$
dimensional space-time border, provided the bulk excitations are suppressed
by a finite energy gap.

Most of the computations of the persistent currents in FQH states 
for finite temperatures have been based so far on the Fermi/Luttinger 
liquid picture. The reason is that
 a more general approach to these quantities was missing.
 In this paper we are trying to
bridge the gap between the general FQH classification scheme based on
unitary CFT and the directly measurable quantities, such as the 
AB magnetization and magnetic susceptibility, 
which express the electro-magnetic properties of the mesoscopic FQH systems.
To this end we shall use as a thermodynamical potential 
the complete chiral CFT  partition function for
the FQH edge states with AB flux  and
a special invariance condition, called the $V$ invariance, representing 
the Laughlin spectral flow\cite{laugh,cz} in the CFT context,
which was introduced by Cappelli and Zemba \cite{cz}.

Typically the  persistent currents in the FQH states contain  
large non-mesoscopic contributions \cite{geller-vignale,michael},
slowly varying with the magnetic field, coming
from states which are deep below the Fermi level.
These contributions cannot be described within the CFT framework 
which takes into account only states close to the Fermi energy
and  therefore the edge states effective CFT is relevant only for 
the computation of the (oscillating) mesoscopic part of the persistent 
currents,
which is proportional to the AB magnetization.
Fortunately, it is possible to measure directly the oscillating part of
the persistent currents due to the
DC Josephson effect in a 2-terminal superconducting ring,
which is known as a Superconducting Quantum Interference Device (SQUID)
\cite{thouless:top,pers-exp}. 
Perhaps, the persistent current and the AB magnetization
are  the only CFT-based prediction of electro-magnetic properties
of the FQH systems  that could be tested directly.
As we shall see in Sect.~\ref{sec:low-T},
these currents are expected to be diamagnetic at very low temperature for 
any FQH state.

Although the FQH states do emerge in highly correlated electron systems 
it was confirmed  both numerically \cite{chakra-pietl-1,chakra-pietl-2} 
and experimentally \cite{pers-exp}   that the electron--electron interaction
does not change significantly the value of the persistent currents. 
On the contrary, it was found that the amplitude of the current is 
strongly reduced by impurities \cite{chakra-pietl-1,chakra-pietl-2}, 
however, since the  persistent currents for a disk FQH sample  are chiral
(for states without  counter-propagating modes on the same edge) 
there should be no impurity 
backscattering,  hence no amplitude reduction from weak 
disorder \cite{geller-loss}. Therefore, the temperature
dependence of the persistent currents in mesoscopic FQH edges
is expected to be universal.
While the persistent currents  naturally express the electric 
properties of the system, it turns out that their universal 
temperature dependence 
 could also give important information
about the neutral structure of the effective field theory for the edge
states \cite{geller-loss-kircz,ino},
and  may be considered as an
ideal tool for distinguishing between different FQH universality classes.
As we shall demonstrate in Sect.~\ref{sec:low-T} and
Sect.~\ref{sec:high-T} the low- and high- temperature decays of
the persistent current's amplitudes crucially depend on the neutral
properties  of the edge states.

 In 1999 the first several FQH states in the so called
 parafermionic (PF) hierarchy in the second Landau level
 with filling factors 
\beq\label{nu_H}
\nu_H:=\frac{n_H}{d_H}= \frac{k}{k+2},
\eeq
introduced by Read and
 Rezayi \cite{rr} have been observed in
 an extremely high-mobility sample \cite{pan} (for FQH states in the second 
Landau level one has to add\cite{rr,cgt2000} $2$ to the filling factor 
in Eq.~(\ref{nu_H})). Originally
  the wave functions of the ground state and excited
 states of these novel FQH liquids 
have been obtained \cite{rr}  as conformal blocks of the well-known
 $\Z_k$ parafermions, however,  in a subsequent work \cite{cgt2000} the
 effective CFT have been expressed as a diagonal affine coset within
 an abelian parent CFT (a multi-component  Luttinger liquid).
 The latter approach allowed to find a much simpler expression for the
 wave functions as well as to interpret the non-abelian
 quasiparticles  as a result of certain projection of neutral
 degrees of freedom.

In this paper we are going to  illustrate the  
 general CFT approach to the chiral persistent currents 
on the  example of the parafermion FQH states in the second Landau 
level\footnote{
We focus on the last occupied Landau level and for the physical measurements
one needs to add also the contributions from the two lowest Landau levels
completely occupied by electrons of both spins.
}. 
Our main result is the observation that all mesoscopic quantities, 
such as the persistent current, AB magnetization and magnetic 
susceptibility, are periodic functions of the AB flux with period exactly one 
flux quantum $h/e$.
Let us  note that the Bloch theorem does not forbid fractional periods, 
such as $1/2$, $1/3$, $1/4$, etc., however, this is always a signal of 
spontaneous breaking of symmetries. Thus, our numerical results rule out 
this possibility for finite temperatures like it should be in view of the 
Mermin--Wagner\cite{mermin-wagner} and Coleman \cite{coleman} theorems.

We shall investigate the temperature dependence of the persistent currents
much above some temperature scale $T_0=\hbar v_F/\pi k_B L$, 
where $v_F$ is the Fermi velocity on the edge and $L$ is the 
edge's circumference, where the current's amplitudes decay 
exponentially with increasing the temperature. 
The corresponding exponents are universal and can be 
used to characterize the universality classes of the FQH states.
For example, for the Laughlin FQH states with $\nu_H=1/(2p+1)$,
 the amplitude of the persistent  current decays according 
to\cite{geller-loss,5-2}
\beq\label{Laugh-high-T}
I_{\max}^{\mathrm{Laugh}}(T) \mathop{\simeq}_{T\gg T_0} \ I_0\  \frac{T}{T_0} \
\exp\left(- \alpha \frac{T}{T_0} \right), \quad \alpha=\frac{1}{\nu_H}
\eeq
(notice that the term $T/T_0$  is missing in Eq.~(95) in 
Ref.~\cite{geller-loss}  which is a misprint). 
For Fermi liquids $\alpha=1$ and any different value is the fingerprint 
of a non-Fermi liquid. A very interesting question
in this context, which was one more motivation for our numerical calculations,
 is \textit{what would be the value of the universal exponent $\alpha$ 
for FQH states  for which the numerator of $\nu_H$ is bigger than $1$?} 
As we shall see in this paper, the answer to this question is not trivial and 
two obvious generalizations, namely
$\alpha=\nu_H^{-1}$ and $\alpha=n_H d_H$ (see Eq.~(\ref{nu_H})), 
are inconsistent with the numerical data for the 
parafermion FQH states.  Our result,  Eq.~(\ref{alpha}), which is fairly 
instructive about the general pattern, 
shows that there is a non-trivial contribution from the neutral sector.
To the best of my knowledge, this is the first generalization of this kind 
which is confirmed numerically.

 The rest of this paper is organized as follows: in
 Sect.~\ref{sec:part} we first describe the relation between the 
Cappelli--Zemba factors, the AB flux  and the Laughlin spectral flow
and then summarize the results from
 Ref.~\cite{cgt2000} about the partition functions 
for the edge states  of the parafermion FQH states for $k=2,3$ and $4$.
In Sect.~\ref{sec:pers} we present the numerical calculations for 
the persistent currents and magnetic susceptibilities at 
fixed temperature,  when the flux is varied within the range  of one 
flux quantum,  as well as for the temperature dependence  of their amplitudes.
 In Sect.~\ref{sec:low-T} we derive the low-temperature
 asymptotics of the persistent currents and susceptibilities
for these states  and try to identify the possible mechanism for the 
amplitude reduction. 
In  Sect.~\ref{sec:high-T} the high-temperature asymptotics is
 computed with the help of the modular S-transformation
 which relates the low- and high-temperature  regimes. 
Another mechanism for the temperature decay of the persistent currents
is suggested for this regime. 
It is shown that  both low- and high- temperature behaviors 
depend crucially on
 the neutral properties  of the FQH system. 
Finally in Sect.~\ref{sec:period} we collect the  arguments  
for our conclusion  that the periods of the 
persistent currents in the parafermion FQH states is exactly one flux 
quantum.
\section{Chiral partition function for the FQH states 
with Aharonov--Bohm flux}
\label{sec:part}
\subsection{Cappelli--Zemba factors  and the Laughlin spectral flow}
\label{sec:CZ}
The partition function $Z(\t,\z)$ for a disk FQH sample with a single edge,
which we shall call  the  \textit{chiral partition function},
computed within the effective CFT for the edge states
is simply the sum of all non-equivalent chiral CFT 
characters\cite{NPB-PF_k}  $\chi_{\l}(\t,\z)$ 
\beqa\label{Z_chi}
Z(\t,\z) &=& \mathop{\mathrm{tr}}_{\ \ \H \ } q^{L_0-c/24}\,
\e^{2\pi i \z J_\mathrm{el}} =
\sum_{\l=1}^N \chi_{\l}(\t,\z), \ \mathrm{where}   \nn
\chi_{\l}(\t,\z) &=&  \mathop{\mathrm{tr}}_{\ \ \H_{\l} \ }
q^{L_0-c/24}\,  \e^{2\pi i \z J_\mathrm{el}}\ \  \mathrm{and} \ \
\H=\mathop{\oplus}_{\lambda=1}^N \H_\lambda.
\eeqa
The number $N$ in Eqs.~(\ref{Z_chi}), called the 
\textit{topological order}\cite{wen-top}, (the number of topologically 
inequivalent quasiparticles with electric charge\cite{fro2000} 
$0\leq Q_{\mathrm{el}} <1$)  
is one of the most important characteristics of the FQH universality classes.
The (Hilbert) space of the edge states $\H$, over which the trace in 
Eq.~(\ref{Z_chi}) is taken, is the direct sum of all 
independent (i.e., irreducible, topologically non-equivalent)
representations  $\H_\lambda$ of the chiral algebra 
(see the Introduction of Ref.~\cite{gaps} for a short description of 
the chiral algebra terminology in the FQH effect).
The effective Hamiltonian for the edge states (in the thermodynamic limit)
is essentially given by \cite{cz}
\beq\label{H_CFT}
H_{\mathrm{CFT}}= \hbar \frac{v_F}{R} \left( L_0 -\frac{c}{24}\right),
\quad L_0=\oint\frac{d\,z}{2\pi i} \ z\, T(z)
\eeq
where $z=\exp[(v_F t-i x)/R]$, $x$ is the coordinate on the edge, $t$ is the 
imaginary time,  $v_F$ is the Fermi velocity on the edge, $R$ is its radius,
 $L_0$   is the zero mode of the Virasoro stress tensor $T(z)$ and $c$ is 
the central charge of the latter \cite{CFT-book}. 
While it has become a wide-spread opinion that the CFT 
Hamiltonian~(\ref{H_CFT}) 
captures only the universal properties of realistic FQH systems  
and eventually 
could not give a physical  description of the latter, 
it has been shown \cite{ctz3} that  the energy spectrum of
the FQH edge excitations, with  Coulomb and generic short-range interactions 
included, contain only logarithmic corrections  to that derived 
from Eq.~(\ref{H_CFT}) which are 
subleading in the thermodynamic limit. 
Therefore, we believe that the CFT approach to the FQH effect could 
give  more information about the structure of the Hilbert space
of the low-laying excitations\cite{ctz3}, the energy spectrum of the FQH 
systems in the thermodynamic limit and even about the energy gap \cite{gaps}.
In particular, it  should not be a surprise
 that  the CFT-based computations \cite{ino,5-2} of the persistent currents
in the Laughlin FQH states give the same results as those based on the 
Luttinger liquid picture \cite{geller-loss}.

The electric charge operator 
$J_\mathrm{el}$ in Eq.~(\ref{Z_chi}), defined as the space integral of the 
charge density on the edge $J_\mathrm{el}(z)=\sqrt{\nu_H} J(z)$, 
proportional to the 
normalized $\widehat{u(1)}$ current $J(z)$ and has the following 
(short-distance) operator product expansion \cite{CFT-book,cz,gaps} 
\[
J_\mathrm{el}=\oint\frac{d\,z}{2\pi i} \ J_\mathrm{el}(z),
\quad 
J_\mathrm{el}(z)J_\mathrm{el}(w)\simeq \frac{\nu_H}{(z-w)^2}.
\]
The modular parameters $q$ and $\z$ entering  Eq.~(\ref{Z_chi}) 
 are generically restricted for an annulus sample by the convergence 
conditions 
\[
q=\e^{2\pi i\t}, \quad 0< \Im\t\sim \frac{1}{k_B T}, \quad \z\in \C
\]
and their real and imaginary parts are related to the inverse temperature, 
 chemical potential and Hall voltage \cite{cz}. 
For the disk FQH sample, however,
both  $\t$ and $\z$ have to be purely imaginary in order for
 the partition function Eq.~(\ref{Z_chi}) be real. Before we establish
the exact relation between the modular parameters and the physical quantities 
for the disk FQH sample we need to recall some of the general properties 
of the corresponding CFTs.
The characters $\chi_\lambda(\t,\z)$ in Eq.~(\ref{Z_chi}),
like those in any rational CFT,  belong to a 
finite dimensional representation of the fermionic subgroup of the 
modular group \cite{cgt2000,fro2000,cz} $PSL(2,\Z)$.
However, in the FQH context there are two more conditions to be 
satisfied \cite{cz}: 
the annulus partition function, which is a bilinear combination of the 
characters $\chi_\lambda(\t,\z)$ and their complex conjugates, 
must be also invariant under the  $U$ and $V$ transformations
of the modular parameters  \cite{cz}. The first one 
simply shifts the parameter $\z$, i.e.,  $U: \  \z\to \z+1$ and the second
acts as follows
\beq\label{V}
V: \ \z\to\z+\t.
\eeq
It is remarkable that the $V$ transformation (\ref{V}) 
exactly represents the Laughlin spectral flow \cite{laugh,cz} in the CFT 
context,
i.e., the mapping between orbitals, corresponding to increasing the orbital 
momentum  by 1 (in the Laughlin FQH states), 
after adiabatically threading the sample with one quantum of 
AB flux. As a result of this flow
a fractional amount of electron charge is transferred between the edges of the
 annulus, which is the basic mechanism inducing the Hall current.

Adding AB flux in the FQH system naturally introduces\cite{NPB-PF_k} twisted 
boundary conditions for the field operators of the electron and the 
quasiparticles and also modifies the Hamiltonian~(\ref{H_CFT}). 
The ultimate effect of this twisting\cite{gt} on the
partition function is that the modular parameter $\z$ is shifted 
\cite{NPB-PF_k} with an amount proportional to the modular parameter $\t$
and the coefficient of proportionality 
has to be identified with the AB flux threading the surface of the FQH sample. 
This coefficient can be computed through the phase that the electron operator 
accumulates when traveling along the FQH edge encircling
the AB flux $\phi$.
In other words, within the rational CFT framework,  
the arbitrary flux ($\phi$) threading procedure is represented by the following
transformation \cite{5-2,NPB-PF_k}
\beq\label{flux}
\z\to\z+\phi \t, \quad \phi\in\R.
\eeq
This generalization of the $V$ transformation~(\ref{V}) 
is obviously  consistent with the  interpretation of the latter  as 
the Laughlin spectral flow.
More intuitively speaking, since the $V$ transformation, Eq.~(\ref{V}),
was originally interpreted as increasing the AB flux by one unit,
the transformation~(\ref{flux})  should mean increasing the flux by 
the  value $\phi$ (in units  $\phi_0$). Note that Eq.~(\ref{flux})
was used to compute the chiral persistent currents in the Laughlin FQH 
states reproducing the well-known result \cite{geller-loss},
 see Appendix B in Ref.~\cite{5-2}.

We shall set $\z=0$ in Eq.~(\ref{flux}), which corresponds to  
zero magnetic field (measured with respect to 
the uniform background magnetic field). Taking into account the interpretation
of Eq.~(\ref{flux}) and choosing the conventional temperature unit $T_0$ 
from Ref.~\cite{geller-loss}
we make the following identification  of the modular parameters
in the context of a disk FQH sample threaded by AB flux $\phi$
\beq\label{mod_param}
\t=i\pi \frac{T_0}{T},\quad \z=\phi\t, \quad 
T_0=\frac{\hbar v_F}{\pi k_B L}, \quad \phi\in \R.
\eeq
In the above equation $T$ is the absolute temperature, $L=2\pi R$ 
is the circumference of the edge,  $k_B$ the Boltzmann 
constant and $\phi$ the magnetic flux threading the FQH disk sample.

The important observation \cite{cz} of Cappelli and Zemba is that the CFT 
partition 
function for the annulus sample is not $V$-invariant alone \cite{cz} since the 
CFT characters $\chi_\lambda$ in Eq.~(\ref{Z_chi}) change their absolute values
after the transformation~(\ref{V}). In order to preserve the norm of the 
characters $\chi_\lambda$, hence restore their $V$-covariance,  
 these authors  introduce special non-holomorphic exponential factors
\beq\label{CZ}
\exp\left(-{\pi}\nu_H\frac{\left(\Im\z\right)^2}{\Im\t}\right),
\eeq
multiplying the characters $\chi_\lambda(\t,\z)$ in Eq.~(\ref{Z_chi}), 
which we shall call the 
Cappelli--Zemba (CZ) factors. In the same way the total chiral partition
function $Z(\t,\z)$ in Eq.~(\ref{Z_chi}) becomes $V$-invariant 
only after multiplying
the characters with the CZ factors~(\ref{CZ}).
One consequence of the $V$-invariance condition is that one has to include 
all twisted sectors which are connected by the action of the spectral 
flow Eq.~(\ref{V}). This restricts the chiral partition function for the disk 
FQH sample to have the ``diagonal'' form like in Eq.~(\ref{Z_chi}).
The physical interpretation of the multiplication with the CZ 
factors~(\ref{CZ})
is adding constant (capacitive) energy to both edges which makes the 
ground state energy independent\cite{cz} of the edge potential $V_0$.
In the next sections we are going to show that the CZ factors play a
fundamental role in the equilibrium thermodynamic phenomena, such as the 
persistent currents and AB magnetization. We believe that they could also 
be crucial in various  transport processes. 
\subsection{Partition function for the  $\Z_k$ parafermion FQH states}
The CFT for the $\Z_k$ parafermions, which is denoted by \cite{cgt2000,gaps}
\beq\label{struct}
\left(\widehat{u(1)}\oplus \PF_k \right)^{\Z_k}, \quad
\PF_k = \frac{\widehat{su(k)}_1\oplus\widehat{su(k)}_1}{ \widehat{su(k)}_2}
\eeq
contains a $u(1)$ factor describing the charge/flux degrees of freedom
and a neutral parafermionic factor which is realized as a diagonal affine coset
\cite{cgt2000}. Both factors are subjected to a $\Z_k$ selection rule called 
the \textit{pairing rule}\cite{cgt2000}, which states that an excitation
\[
:\e^{i\frac{\l}{\sqrt{k(k+2)}}\varphi(z)}: \otimes \ \Phi(z)
\] of 
the $\PF_k$ model with labels $(\l,\Phi)$, where 
$\varphi$  is the $u(1)$ chiral boson,  $\l$ is the corresponding $u(1)$ 
charge label and $\Phi$ is a parafermion (primary) field 
\cite{cgt2000,CFT-book}, is  legitimate only if
\[
P\left[ \Phi\right]\equiv \l \ \mod \ k,
\]
with $P$ being the parafermion $\Z_k$ charge.
The independent characters $\chi_{l,\rho}$ of the chiral CFT
(whose number gives the topological order) for the 
parafermion FQH states in the second Landau level  \cite{cgt2000}
are labeled by two integer numbers  $l \mod k+2$ (symmetric around $l=0$) 
and $\rho=0,\ldots,k-1$ with the restriction $\rho\geq l-\rho \mod k$
\beq\label{chi_PF}
\chi_{l,\rho}(\t,\z) =
\sum_{s=0}^{k-1} K_{l+s(k+2)}(\t,k\z;k(k+2))
\ch\left(\L_{l-\rho+s} +\L_{\rho+s}\right)(\t)
\eeq
and the complete chiral partition function is defined as
\beq\label{Z_PF}
Z_{k}(\t,\z) =
\e^{-{\pi}\nu_H\frac{\left(\Im\z\right)^2}{\Im\t}}
\sum_{l  \mod  k+2} \quad
\sum_{\rho \geq l-\rho \mod k } \chi_{l,\rho}(\t,\z),
\eeq
where we have written the CZ factors~(\ref{CZ}) explicitly.
The rational torus partition functions \cite{CFT-book} $K_l$
represent the contribution of the charged sector
\beq\label{K}
K_l(\t,\zeta;m)=\frac{1}{\eta(\t)} \sum_{n\in\Z}
q^{\frac{m}{2}\left(n+\frac{l}{m}\right)^2} 
\e^{2\pi i \zeta \left(n+\frac{l}{m}\right)},
\eeq
where $\eta$ is the Dedekind function \cite{CFT-book}
\beq\label{Dede}
\eta(\t) = q^{\frac{1}{24}} \prod_{n=1}^\infty \left(1-q^n\right),
\eeq
while the neutral characters $\ch(\L_\mu+\L_\rho)$ 
which are labeled\cite{cgt2000} 
by the $\widehat{su(k)_2}$ weights $\L_\mu+\L_\rho$
with $0\leq\mu\leq\rho\leq k-1$, are expressed
in terms of the so called universal chiral partition functions
\cite{cgt2000} $\ch^Q_\s$  with the identification
$Q=\rho$, $\s = \rho-\mu$
\beq\label{ch}
\ch^Q_{\s}(\t)=q^{\D_\s-\frac{c_{\PF}}{24}}
\sum_{
\mathop{n_1,n_2,\ldots,n_{k-1}=0}\limits_{
	 [\underline{n}]	\equiv Q \mod k}}^\infty
	\frac{q^{\underline{n}\cdot C^{-1}\cdot (\underline{n}- \L_{\s} ) }}
{ (q)_{n_1}\cdots (q)_{n_{k-1}}},
\quad Q\geq \s.
\eeq
In the above equation, $\underline{n} =(n_1,n_2,\ldots,n_{k-1})$ is
a $k-1$ component vector with non-negative entries,
 $[\underline{n}]=\sum\limits_{i=0}^{k-1} i \, n_i$ is its $k$-ality,
\beq\label{Delta_s}
\D_\s= \D^{\PF}\left(\L_0+\L_\s \right) = \frac{\s(k-\s)}{2k(k+2)}
\eeq
 is the  CFT dimension of the  $\PF_k$ primary field labeled 
by\cite{cgt2000} $\L_0+\L_\s$, $c_{\PF} =\frac{2(k-1)}{k+2}$
is the parafermion central charge,
$ (q)_n=\prod\limits_{j=1}^n (1-q^j)$
and $C^{-1}$ is the inverse of the $su(k)$ Cartan matrix.

In order to compute the partition functions with AB flux, we have to 
apply the flux transformation
(\ref{flux}) to the partition function (\ref{Z_PF_0}).
To this end we shall use the following important 
property\footnote{moving the flux dependence
into the indices of the $K$-functions is technically convenient since it
guarantees the reality of the disk sample partition functions when 
computed numerically
for all values of $\phi$} of the $K$-functions
\beqa\label{K-flux}
& \e^{-{\pi}\nu_H\frac{\left(\Im(\z+\phi\t)\right)^2}{\Im\t}}
K_l(\t,n_H(\z+\phi\t);n_H d_H)) =  & \nn
& = \e^{-{\pi}\nu_H\frac{\left(\Im\z\right)^2}{\Im\t}}
K_{l+n_H\phi}(\t,n_H\z;n_H d_H) &
\eeqa
applied for $\z=0$. 
\section{Persistent currents and magnetic susceptibilities}
\label{sec:pers}
The equilibrium thermodynamic \textit{chiral persistent current} in 
the disk FQH system can be computed directly from the CFT partition 
function (\ref{Z_chi}) as follows
\beq\label{pers}
I(T,\phi)=\left(\frac{e}{h}\right) k_B T \frac{\partial}{\partial \phi}
\ln Z(\t,\phi\t),
\eeq
where the temperature $T$ and the AB flux $\phi$ are related to the modular 
parameters $\t$ and $\z$ according to Eq.~(\ref{mod_param}).
More explicit formulas for the chiral partition functions (\ref{Z_PF})
and (\ref{chi_PF}) for the 
parafermion FQH states, before and after threading the sample with 
arbitrary AB flux, can be found in Ref.~\cite{5-2,NPB-PF_k}.

\begin{figure}[htb]
\centering
\caption{Persistent currents in the $k=2,3$ and $4$ parafermion FQH states,
as functions of the magnetic flux within one period,
computed numerically for $T/T_0=0.1$.
The flux is measured in units $h/e$ and the current's unit is
$e v_F/L$.
The period is 1 flux quantum for all states and the amplitudes are
$I_2^{\max}=0.207$,  $I_3^{\max}=0.245$ and $I_4^{\max}=0.268$.
\label{fig:periods}}
\epsfig{file=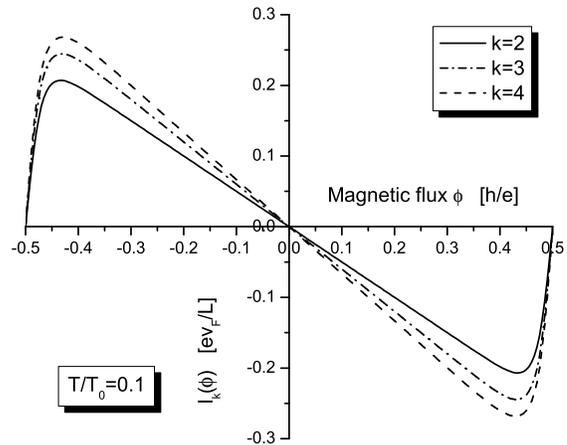,height=7cm}
\end{figure}

We have computed numerically the persistent currents using the 
definition Eq.~(\ref{pers}) and the explicit formulas 
Eqs.~(\ref{Z_PF}), (\ref{chi_PF}), (\ref{K}) and (\ref{ch})
for the characters of the parafermion FQH states.
All computations have been performed with unlimited 
 number  of terms for 
the partition functions (\ref{K}) and finite number  
($10^k$, with $k=2,3$ and $4$ respectively) of terms 
for the partition functions (\ref{ch}).
The plots of the persistent currents profiles in
the $\PF_k$ states  with  $k=2$, $3$ and $4$ and AB flux in the range
$-1/2 \leq \phi \leq 1/2$ for temperature $T/T_0=0.1$ 
are given on Fig.~\ref{fig:periods}.
According to the discussion in Sect.~\ref{sec:CZ}, 
the partition function~(\ref{Z_PF})
constructed for the  $\PF_k$ states must be invariant under the
$V$ transformation  Eq.~(\ref{V}), which
implies that the corresponding  chiral persistent current should be
periodic in $\phi$ with period at most  $1$. 
The plots in Fig.~\ref{fig:periods}, indicate that
\textit{all currents are periodic in $\phi$ with a period exactly one quantum 
of flux ($h/e$)}.
We  do not see  anomalous oscillations with 
fractional flux periodicity\footnote{under certain conditions the period 
of the persistent currents for the paired states ($k=2$) 
could be shown\cite{ino:comm} to be $1/2$} 
of the persistent currents \cite{ino,ino2,kiryu}
 in the $k=2,3,4$ $\PF_k$ states for all temperatures
$0.03\leq T/T_0 \leq 14$ that we could access numerically.
This is in perfect agreement with  the Bloch theorem, known also as
the Byers--Yang theorem \cite{byers-yang} in the context of
superconductors.
Notice that the persistent currents for the Read--Rezayi FQH states  
 have been shown in Ref.~\cite{kiryu}
to have anomalous oscillations  for $M=0$ with periods smaller than $1$, 
however we argue that 
this case is physically irrelevant and has only been used in
Ref.~\cite{rr} as a technical tool for investigating the wave functions
for the $\PF_k$ states. One of the objectives of this paper was to show
that the periods of the persistent currents for the $\PF_k$ states
are exactly $1$ for all non-zero temperatures.

\begin{figure}[htb]
\centering
\caption{Magnetic susceptibilities in the $k=2,3$ and $4$ parafermion 
FQH states,
as functions of the magnetic flux within one period,
computed numerically for $T/T_0=0.1$.
The flux is measured in units $h/e$ and the 
susceptibility  unit is 
$v_F\frac{\mu_0}{4\pi}\frac{e^2}{h} \frac{\pi R^2}{wd}$, where $w$ and $d$ 
are respectively the finite width of the ring and the finite thickness of 
the 2 dimensional electron layer.
The period is 1 flux quantum for all states.
\label{fig:susc-per}}
\epsfig{file=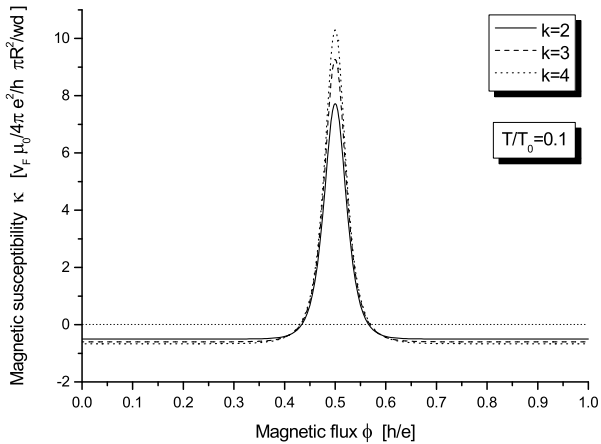,height=7cm}
\end{figure}

It is more realistic to consider the edge of a  FQH disk sample 
as a very thin ring (with inner radius $R_{\mathrm{in}}$ and outer one 
$R_{\mathrm{out}}$) rather than just a circle, in which case $R$ is
the average radius of the ring $R=(R_{\mathrm{in}}+R_{\mathrm{out}})/2$. 
Under the assumption that the width $w=R_{\mathrm{out}}-R_{\mathrm{in}}$ 
of the ring and the thickness
$d$  of the physical 
2-dimensional electron layer are much smaller than the  ring's 
circumference $L$, 
the magnetization of the FQH ring, like all thermodynamic quantities, 
can be expressed as a function of the magnetic flux 
(not of the magnetic field alone), and becomes 
simply proportional to the persistent current, i.e.,
\[
M(T,\phi)=-\frac{1}{\mathrm{Vol}} \frac{\partial F(T,\phi)}{\partial B} =
\frac{L}{4\pi w d}\ I(T,\phi),
\]
where $\mathrm{Vol}=2\pi R w d$ is the volume of the physical mesoscopic ring
(with finite $w$ and $d$) and we have 
used $\partial /\partial B = \pi R^2\phi_0^{-1}\, \partial /\partial\phi$.
Furthermore, the (isothermal) magnetic susceptibility is   
proportional to the derivative of the persistent current with respect to 
the flux
\beq\label{susc}
\kappa = \left. \frac{\partial M}{\partial H} \right\vert_{H=0}=
\mu_0 \frac{e}{h} \pi R^2  
\left.\frac{\partial M}{\partial \phi} \right\vert_{\phi=0}.
\eeq
The magnetic susceptibilities for the parafermion FQH states with $k=2,3$ 
and $4$ have also been computed numerically and they
also show periodicity with the flux with a period of  one flux quantum
exactly for all  temperatures  
$0.03 \leq T/T_0 \leq 14$. 
The plots of the susceptibilities for temperature $T/T_0=0.1$ as 
functions of the flux within the range $0\leq \phi \leq 1$  are given 
on Fig.~\ref{fig:susc-per}.
We have chosen  a different flux 
range in Fig.~\ref{fig:susc-per} in order to show the entire pick 
at $\phi=1/2$.

\begin{figure}[htb]
\centering
\caption{Temperature decay of the persistent current's amplitudes
in the $k=2,3$ and $4$ parafermion FQH states (without the contribution from
the two $\nu=1$ Landau levels with opposite spins)  computed numerically
in units $e v_F/L$, for temperatures measured in units of $T_0$.
The zero temperature amplitudes in these units are $\nu_k/2$, i.e.,
$1/4,3/10$ and $1/3$ respectively.
\label{fig:decay}}
\epsfig{file=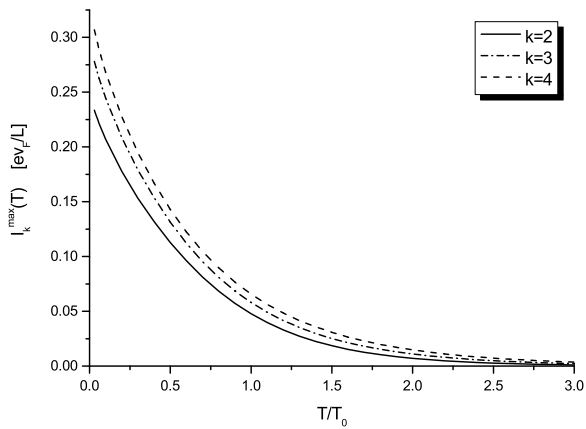,height=6.5cm}
\end{figure}

The amplitudes of the persistent currents in the parafermion FQH
states with $k=2$, $3$ and $4$, according to our computations,  
decay exponentially with temperature, as shown in Fig.~\ref{fig:decay}. 
Notice that, in order to compute these amplitudes  we 
first analytically differentiate the logarithm of the partition function
(\ref{Z_PF})  with respect to $\phi$, according to Eq.~(\ref{pers}), 
to get the persistent current $I(T,\phi)$, then differentiate analytically 
again with respect to $\phi$  
and finally solve numerically $\partial I(T,\phi)/ \partial\phi=0$  
in the interval $-1/2 < \phi <0$ to find the position of the maximum.
The positions $\phi_{\max}$ of the maximums of the persistent currents, 
for $k=2$, $3$ and $4$ respectively,  have been also plotted as functions of 
temperature on Fig.~\ref{fig:phi_max} and a more detailed discussion of 
this issue will be given in Sect.~\ref{sec:period}.
The same three curves of Fig.~\ref{fig:decay}, however within the full 
computational range of temperatures $0.03\leq T/T_0 \leq 14$,  
have been also plotted in a 
(semi-) logarithmic plot on Fig.~\ref{fig:log-decay}; the plots on 
Fig.~\ref{fig:log-decay} look 
almost linear, which means almost exponential decays for the amplitudes 
themselves.
Each of the three curves on Fig.~\ref{fig:log-decay} contains  147 points 
and each point would need about 280 seconds (on average) to compute using 
MAPLE-8 on a 2 GHz class PC.
\begin{figure}[htb]
\centering
\caption{Temperature decay of the (zero-field) magnetic susceptibility 
in the $k=2,3$ and $4$ parafermion FQH states  computed numerically
in dimensionless units 
$v_F \frac{\mu_0}{4\pi} \frac{e^2}{h} \frac{\pi R^2}{wd}$, 
where $w$ and $d$ are respectively  the finite width and thickness of a 
realistic 2D electron layer.
The zero temperature susceptibilities in these units are exactly 
$-\nu_k$, i.e., $-1/2$, $-3/5$ and $-2/3$ respectively.
\label{fig:susc-decay}}
\epsfig{file=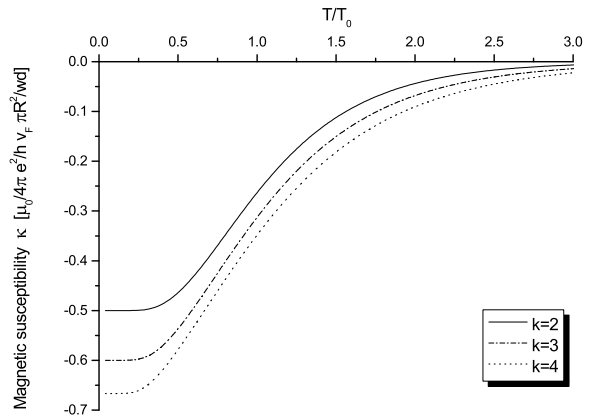,height=6.5cm}
\end{figure}

The magnetic susceptibilities~(\ref{susc}) also decay  very fast with 
temperature as shown on Fig.~\ref{fig:susc-decay}, for the $k=2$, $3$ 
and $4$ $\PF_k$ states.
Moreover, since the susceptibility is a derivative of the persistent 
current, its temperature behavior determines the decay-type of the amplitude
of the persistent current. It could be seen from Fig.~\ref{fig:susc-decay}
that there are two different temperature regimes in which the persistent 
currents' amplitudes decay in different ways. 
Note that the function $\kappa(T)$ has an inflection point close
to $T_0$ which can be interpreted as the crossover temperature 
between the two regimes. As we shall see in more detail 
in Sects.~\ref{sec:low-T} and \ref{sec:high-T}, there are indeed 
two different mechanisms for the temperature decay of the amplitude 
(cf. Ref.~\cite{pers-exp}).
The first one,  for $T\ll T_0$  is due to mixing of
contribution of levels in an energy interval $k_B T$
which reduces the current since adjacent  levels give opposite contributions
 \cite{pers-exp}. The energy scale for this mixing is set  \cite{pers-exp} 
by the  energy gap. 
The second mechanism, which is essential for the current at temperatures 
higher than $T_0$, is a thermal smearing due to
the  reduction of the phase coherence length \cite{pers-exp}
 $L_{\varphi}$ which makes the current vanish exponentially in 
$L/L_{\varphi}$. A more detailed discussion of this point will be given in 
Sect.~\ref{sec:high-T}.
\section{Low-temperature regime: $T\ll T_0$}
\label{sec:low-T}
In this Section we are going to find analytic  expressions
for the persistent current's amplitudes and magnetic susceptibilities
for the $\PF_k$ states in the  limit $T\to 0$. 
Our low-temperature analysis would be valid  for a class of FQH states 
which is larger than the parafermion states in the second Landau level.
It would be based on the natural assumption that 
\textit{the CFT dimension of the 
fundamental quasihole is the minimal one allowed in the CFT}, 
which is of course fulfilled in the $\PF_k$ states \cite{cgt2000,gaps}.

The low-temperature limit $T\to 0$ corresponds to the trivial
limit $q\to 0$.
Therefore we can keep only the first terms in the disk partition
function (\ref{Z_PF})  which, under the above mentioned assumption, 
 come from the vacuum character and from those with the minimal 
non-trivial  CFT dimension,
i.e., from the sectors containing one quasiparticle (qp) or one 
quasihole (qh)  so that
\beqa\label{Z_PF_0}
&Z(\t,\z) \mathop{\simeq}\limits_{\ T\ll T_0 \ }
	\e^{-{\pi}\nu_H\frac{\left(\Im\z\right)^2}{\Im\t}}
\Bigl( K_{0}(\t,n_H\z;n_H d_H)\ch_0(\t)+ &\nn
& K_{1}(\t,n_H\z;n_H d_H)\ch_\qh(\t) +  K_{-1}(\t,n_H\z;n_H d_H)\ch_\qp(\t)
\Bigr), \quad &
\eeqa
where $\nu_H$, $n_H$ and $d_H$ are defined in Eq.~(\ref{nu_H})
and the exponent in front of the parentheses is the CZ factor~(\ref{CZ}).
Here we have denoted by $\ch_\qh=\ch(\L_0+\L_1)$ and 
$\ch_\qp=\ch(\L_0+\L_{k-1})$ the coset characters corresponding
to the neutral components of the quasiparticle and quasihole
in the $\PF_k$ states.

Since, in order to compute the persistent current,  we intend to 
differentiate with respect to $\phi$ according to Eq.~(\ref{pers})),
 we shall skip the $\phi$-independent factors  
containing the $\eta$-function and the central charge, which after taking 
the logarithm will give rise to additive flux-independent terms that will 
drop out after differentiation.
Applying Eq.~(\ref{K-flux}) in order to incorporate the AB flux and 
keeping only the first terms in Eq.~(\ref{Z_PF_0})
 we can write\footnote{after
including the flux the indices of the $K$ functions are modified so
that e.g., $K_{2+k\phi}$ may dominate over $K_{-1+k\phi}$ for
$-1/2 < \phi<0$ in the $T\to 0$ limit. However, in this case both
functions can be neglected when compared to $K_{1+k\phi}$ since
the latter gives the smallest minimal CFT dimension }
\beqa\label{Z_0}
& Z(T,\phi)  \mathop{\simeq}\limits_{\ T\ll T_0 \ }
q^{\frac{\nu_H}{2}\phi^2} \left\{ 1+ q^{\D_\qh}
\left( q^{Q_\qh\phi}+q^{-Q_\qh\phi}\right)\right\}  = &\nn
& =  \e^{- \pi^2 \frac{T_0}{T}\nu_H\phi^2}
\left\{ 1+ 2 \e^{-\frac{\widetilde{\eps}_\qh}{k_B T}}
\cosh\left(2 \pi^2 \frac{T_0}{T}Q_\qh\phi\right)    \right\}  &
\eeqa
where
\beq\label{proper}
\widetilde{\eps}_{\qh}= 2 \pi^2 k_B T_0 \D_\qh, \quad
 \D_\qh=\frac{1}{2n_H d_H} +\D^{(0)}_\qh ,\quad
 Q_\qh=\frac{1}{k+2}.
 \eeq
In the above equation  $\widetilde{\eps}_{\qh}$ is the effective
proper quasihole energy \cite{morf:proper,geller-loss}
(defined at fixed density),   $\D_\qh$ is the total
CFT dimension of the quasihole (with  $\D^{(0)}_\qh$  being its
neutral component)  and $Q_\qh$ is the quasihole's electric charge.
The neutral CFT dimension $\D^{(0)}_\qh=(k-1)/(2k(k+2))$ of the quasihole 
for the $\PF_k$ states is obtained from EQ.~(\ref{Delta_s}) for $\s=1$.
In the derivation of Eq.~(\ref{Z_0}) we have kept only the leading terms 
in the (neutral) coset characters
$\ch_0(\t)\simeq q^0$, $\ch_\qp(\t)=\ch_\qh(\t)\simeq q^{\D^{(0)}_\qh}$
for $q\to 0$ (again dropping the central charge term).
Notice that the partition function (\ref{Z_0})
is finite in this approximation \cite{NPB-PF_k}.

Next, we take the logarithm of Eq.~(\ref{Z_0}) and  differentiate 
with respect to $\phi$ like in Eq.~(\ref{pers}) to get the following 
expression for the  persistent current for $|\phi| < 1/2$
\beq\label{I-low}
I(T,\phi)  \  \mathop{\simeq}_{T\ll T_0} \
\frac{e v_F}{L}
\left\{ - \nu_H \phi +2 Q_\qh \e^{-\frac{\widetilde{\eps}_\qh}{k_B T}}
\sinh\left( 2 \pi^2 \frac{T_0}{T}Q_\qh\phi\right)    \right\}.
\eeq
\subsection{The persistent current's amplitude for $T=0$}
\label{sec:I_0}
For $T=0$ the second term in the curly brackets in Eq.~(\ref{I-low})
vanishes expressing the fact that for zero temperature the free energy is
determined by the ground state energy. Then the persistent current is 
a linear function of the flux 
\beq\label{I_0}
I(0,\phi)= - \nu_H\frac{e v_F}{L}  \, \phi \quad \mathrm{for}\quad
-\frac{1}{2}< \phi < \frac{1}{2}
\eeq
and riches its maximum 
\[
  I_{\max}=\frac{1}{2}\, \nu_H\, \frac{ev_F}{L}  \quad \mathrm{for} \ T=0 .
\]
at $\phi=-1/2$. By periodicity, the current Eq.~(\ref{I_0}) could be continued 
to any value of the magnetic flux, which gives rise to the so called 
saw-tooth curve.
The peak-to-peak value of the amplitude $ev_F/L$ for the mesoscopic 
sample of Ref.~\cite{pers-exp},  for which $v_F\simeq 2.6\times 10^5$~m/s 
and  $R\simeq 2.7 \mu$m,  is approximately $5$~nA.

The zero-field magnetic susceptibility (see Eq.~(\ref{kappa_T}) below) 
at zero temperature  
\beq\label{kappa_0}
\kappa(0)= -\nu_H\frac{\mu_0}{4\pi}\left(\frac{\pi R^2}{wd}\right)
\frac{e^2}{h} v_F , \quad w\ll R, \quad d \ll R
\eeq
has absolute value of the order of $10^{-4}$ for the sample of  
Ref.~\cite{pers-exp}. \\

\noindent
\textbf{Remark 1.}
\textit{The CZ factors (\ref{CZ}) completely determine the zero-temperature
behavior of the FQH system. In particular, they modify the ground
state energy in presence of AB flux and therefore fix the
zero-temperature behavior of the persistent current and magnetic 
susceptibilities.} \\

We stress that Eq.~(\ref{I_0}) and Eq.~(\ref{kappa_0}) are derived 
entirely from the CZ factors and are therefore  valid for all FQH states. 
They clearly show that 
\textit{the persistent currents in mesoscopic FQH edges are diamagnetic
with respect to the AB flux} at zero temperature. 
Our analysis for $T>0$ showed that the persistent
 currents in the parafermion FQH states remain diamagnetic for all 
numerically accessible temperatures $0.03 \leq T/T_0 \leq 14$, 
as illustrated on Fig.~(\ref{fig:periods}), 
Fig.~(\ref{fig:susc-per}) and Fig.~(\ref{fig:susc-decay}). 
A similar  phenomenon has been recently observed  
in a series of experiments with small numbers of metallic mesoscopic 
rings \cite{diamag}. A natural explanation of this result could 
be given by the Lenz rule for the Faraday induction: 
for polarized FQH states, where the magnetic degrees of 
freedom are frozen by the huge background magnetic field corresponding to a 
FQH plateau, the only response of the FQH system to the change in the flux
is the magnetization persistent current, which creates magnetic field that is
always opposite to the direction of the changes of the magnetic field that 
have created the current.
On the other hand, the magnetic flux $\phi$ entering 
Eq.~(\ref{mod_param}) is actually the difference between the total 
magnetic flux 
and that of the background magnetic field. Since the SQUID detectors 
measure exactly the oscillating part of the persistent current, when the 
flux $\phi$ is changed adiabatically, we expect  that its magnetization
should  always be opposite to the direction of the changes of the flux. This 
may explain why the derivative of the persistent current is  negative
at $\phi=0$.  In FQH states that are not completely polarized 
an eventual paramagnetic component could  be dominant.\\

\noindent
\textbf{Remark 2.} \textit{In principle the the persistent currents  in
the FQH states are paramagnetic for
even parity  and diamagnetic for odd parity of the number of electrons
in the mesoscopic ring
\cite{loss-parity,loss-goldbart,kim}. 
However, the parity of the electron number is
not well-defined in the experimental setup of Ref.~\cite{pers-exp}
so that one has to average over the electron parity.
For example the average  of the two currents in Eq. (14) in 
Ref.~\cite{kim} gives $1/2 \,  ev_F/L$ which
  coincides with the CFT prediction for the 
persistent currents in the thermodynamic limit.}\\

Another interesting analogy can be made with the amplitude of the 
current~(\ref{I_0}). It is known \cite{altland} that the zero temperature of 
the (disorder and sample averaged) persistent current for ballistic systems 
is proportional to the Thouless energy
\[
\vert I(\phi)\vert =\frac{e}{h} E_T \phi, \quad \mathrm{for}\quad T=0
\]
and $\phi$ measured in units of $\phi_0$. 
On the other hand the ratio of the Thouless energy and the level spacing 
$\D=\hbar \, 2\pi v_F/L$, which  is known as the Thouless number $g$,
gives the dimensionless conductance. 
In our case this number coincides with the  dimensionless Hall conductance 
\[
E_T=\nu_H \hbar\, \frac{2\pi v_F}{L},\quad  g=\frac{E_T}{\D}=\nu_H.
\]
\subsection{The persistent current's amplitudes for $T>0$}
According to Theorem 3 in Ref.~\cite{byers-yang} the partition function
of the system should be an even periodic function of the flux $\phi$ with
period $1$. Therefore the persistent current (\ref{pers}) must be
an odd function with period $1$ so that it would be enough to
consider the half interval $-1/2 < \phi < 0$ and continue it as an odd
function  $I(\phi)=- I(-\phi)$. 
Writing the $\sinh$ in Eq.~(\ref{I-low}) into its
 exponential form and ignoring $\exp(2\pi^2 T_0 Q_\qh \phi/T)$ which vanishes
 in the limit $T\to 0$ since $\phi$ is negative,  we get
 \beq\label{I-low-T}
 I(T,\phi) \  \mathop{\simeq}_{T\ll T_0} \
\frac{e v_F}{L}
\left\{ - \nu_H \phi - Q_\qh \e^{-\frac{\widetilde{\eps}_\qh}{k_B T}}
\e^{-2 \pi^2 \frac{T_0}{T}Q_\qh\phi}    \right\}.
 \eeq
In order to find the local maximum of the function $I(T,\phi)$
for fixed $T$ we have to solve the stationarity equation
$\partial I(T,\phi)/\partial \phi =0$ and its unique solution 
for  $-1/2 < \phi \leq 0$ is
\beq\label{phi_max}
\phi_{\max}(T) \  \mathop{\simeq}_{T\ll T_0} \
 - \frac{\D_\qh}{Q_\qh} +\frac{1}{2\pi^2 Q_\qh}\frac{T}{T_0}
\ln\left( \frac{2\pi^2Q_\qh^2}{\nu_H}\frac{T_0}{T}\right) .
\eeq
Substituting Eq.~(\ref{phi_max}) into Eq.~(\ref{I-low}) and
using the charge--statistics relation\cite{gaps,NPB-PF_k}
 $2\D_\qh=Q_\qh $,  as well as the value of the
proper quasihole energy Eq.~(\ref{proper}), 
we get the following low-temperature asymptotic expression for the amplitude
\beq\label{I_max2}
I_{\max}(T)\  \mathop{\simeq}\limits_{T\ll T_0} \ \nu_H   \frac{e v_F}{L}
\left\{ \frac{1}{2} -\frac{k_B T}{2 \widetilde{\eps}_\qh}
  \left[1+\ln\left( \frac{1}{n_H}\frac{2 \widetilde{\eps}_\qh}{k_B T}\right) 
\right]
	\right\}.
\eeq
Eq.~(\ref{I_max2}) tells us that the persistent current's amplitude 
in any FQH state, satisfying the condition formulated in the beginning of 
Sect.~\ref{sec:low-T},  decays logarithmically (for $T\ll T_0$)
with increasing the temperature due to the probability for thermal activation
of quasiparticle--quasihole pairs, which flowing to the opposite 
edges reduce the corresponding edge charges hence the radial electric
field that is responsible for the appearance of the (axial) persistent
current.  Indeed, the negative sign in front of the proper quasihole energy in
Eq.~(\ref{I_max2}) implies that the thermally activated quasiparticles
contribute to the current in the  direction opposite to that 
of the ground state's current, i.e., reducing the amplitude 
as stated in  Ref.~\cite{pers-exp}. 
The characteristic energy for this effect  is
 twice the proper quasihole energy $\widetilde{\eps}_\qh$, which plays the
 role of the energy gap in the absence of disorder.

\begin{figure}[htb]
\centering
\caption{Low-temperature asymptotics of the persistent current's amplitudes
in the $k=2,3$ and $4$ parafermion FQH states,  computed numerically
($k=2$ N, $3$ N, $4$ N)  and analytically ($k=2$ A, $3$ A, $4$ A) 
from Eq.~(\ref{I_max2}), in units $e v_F/L$.
The zero temperature amplitudes  are $\nu_k/2$ in these unit, i.e.,
$1/4,3/10$ and $1/3$,  respectively.
\label{fig:num-asy}}
\epsfig{file=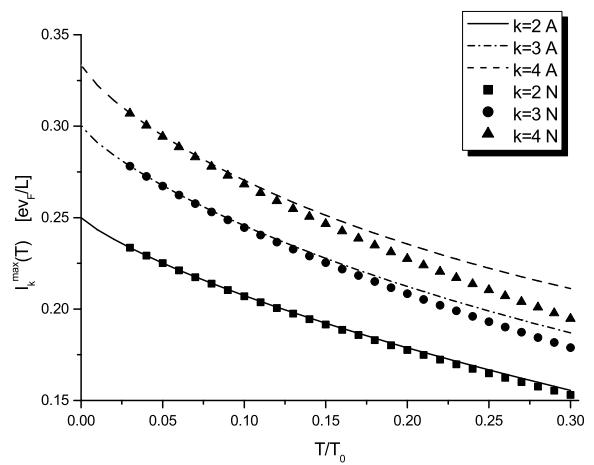,height=7cm}
\end{figure}

The low-temperature asymptotic expressions (\ref{I_max2}) for the amplitude of 
the persistent currents in the $\PF_k$ states with $k=2$, $3$ and $4$ 
are compared in Fig.~\ref{fig:num-asy} to the corresponding exact 
quantities computed numerically.
This figure shows that the asymptotic formula
Eq.~(\ref{I_max2}), for the  amplitudes of the persistent 
currents at low-temperature,  is an excellent approximation for $T/T_0<0.1$.
In addition, the asymptotic expression Eq.~(\ref{I_max2})  allows us
to compute the derivative of the amplitude's temperature decay for $T\to 0$.
While it may seem reasonable to expect that the amplitudes become temperature
independent for $T\to 0$,  Eq.~(\ref{I_max2}) shows that
the derivative $d\, I_{\max}(T)/d\,T$ diverges logarithmically at zero 
temperature.

Applying the definition (\ref{susc}) to the  low-temperature asymptotic 
form (\ref{I-low-T}) of the persistent current we obtain the following 
low-temperature expression for the magnetic susceptibility (for $\phi=0$)
\beq\label{kappa_T}
\kappa(T)   \mathop{\simeq}\limits_{T\ll T_0} 
v_F\frac{\mu_0}{4\pi}\frac{e^2}{h}\frac{\pi R^2}{wd} 
\left\{ -\nu_H  +  2\pi^2 Q_\qh^2 \frac{T_0}{T}
\exp\left(-\frac{\widetilde{\eps}{_\qh}}{k_B T}\right)\right\},
\eeq
where $w$ and $d$ are respectively the finite width of the ring and the 
finite thickness of the 2D electron layer.
The first term in the curly brackets in the above equation gives  the zero 
temperature susceptibility~(\ref{kappa_0}) and its negative sign expresses 
the fact that the persistent currents in the mesoscopic FQH edges are 
diamagnetic, which is in agreement with the discussion in Sect.~\ref{sec:I_0}. 
The second term expresses the reduction of the 
magnitude of the (zero field) susceptibility  $\kappa$ 
due to thermal activation of quasiparticles.
We see from Fig.~(\ref{fig:susc-decay}) that the zero-field 
magnetic susceptibility remains negative for all temperatures and
approaches zero from below for $T\gg T_0$. 
\section{High-temperature regime: $T\gg T_0$}
\label{sec:high-T}
\begin{figure}[htb]
\centering
\caption{Logarithmic plot of the temperature dependence of the persistent 
current's amplitudes 
in the $k=2,3$ and $4$ parafermion FQH states
computed numerically for temperatures in the range  
$0.03\leq T/T_0 \leq 14$ \label{fig:log-decay}}
\epsfig{file=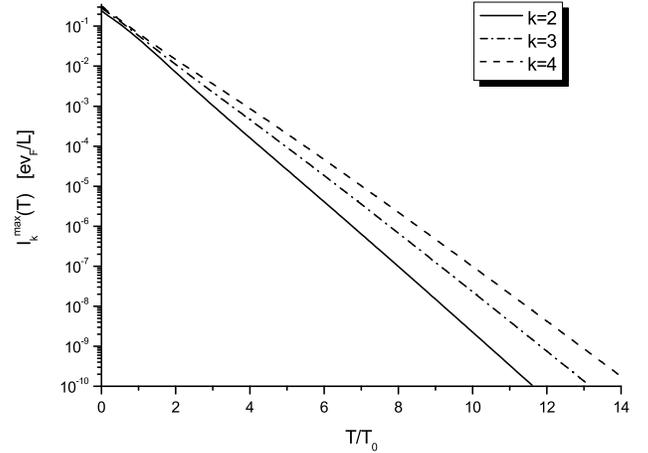,height=7cm}
\end{figure}

In this section we shall find asymptotic expressions for the persistent 
currents in the $k=2$, $3$ and $4$ parafermion FQH states at temperatures
higher than $T_0$. The logarithmic plot of the persistent current's amplitudes
for the range $0.03\leq T/T_0 \leq 14$, which is  
given  on Fig.~\ref{fig:log-decay}, shows that for $T>T_0$
the logarithmic amplitudes decrease almost linearly with $T$, which means 
that the amplitudes themselves  decrease exponentially.
Before we derive the high-temperature asymptotics,
based on the CFT partition functions for the edge states,
let us first announce the result: for 
$T\gg T_0$ the persistent currents 
amplitudes for the  $k=2,3$ and $4$ parafermion FQH states decay 
exponentially with temperature according to 
 \beq\label{high-T}
\bar{I}_{k}(T) \ \mathop{\simeq}_{T\gg T_0} \ I^0_k \left(\frac{T}{T_0}\right)
\exp\left(-\alpha_k \frac{T}{T_0}\right)
 \eeq
 where $I^0_k$ are constants and  the universal exponents $\alpha_k$
can be written explicitly in the form
 \beq\label{alpha}
 \alpha_k = \frac{1}{\nu_H}+2\D^{\PF_k}\left(\L_0+\L_2\right)= 
\frac{k+6}{k+2},
 \eeq
where $\nu_H$ is the filling factor~(\ref{nu_H}) and 
$\D^{\PF_k}\left(\L_0+\L_2\right)$ is computed from 
Eq.~(\ref{Delta_s}) for $\s=2$.\\

\noindent
\textbf{Remark 3.}
\textit{
 It is worth-stressing that the universal exponents (\ref{alpha}) for
the $\PF_k$ states are  not simply equal to the inverse of the
 filling factors like it is in the Laughlin states
\cite{geller-loss,geller-loss-kircz,geller-loss-kircz-1} 
(see Eq.~(\ref{Laugh-high-T})).
 There are  crucial  contributions coming  from the neutral sectors which are  
described here by the  $\Z_k$ parafermion cosets.
}\\

Now, let us describe in more detail our analytic computations for
the high-temperature regime of the persistent currents in the 
$\PF_k$ states. 
The high-temperature limit $T\to \infty$
(unlike the low-temperature one) is rather non-trivial
since the modular parameter $q=\exp(-2\pi^2 T_0/T) \to 1$ is going outside 
of the convergence interval for the partition functions and therefore cannot be
taken directly. In our case however, one could use the advantage that the
complete chiral partition function is constructed as a sum of rational 
CFT characters,
which are covariant under the modular 
$S$-transformation\cite{CFT-book,cgt2000}, i.e., 
\[
\chi_{\l}(\t,\z)= \sum_{\l'=1}^N S_{\l\l'} \
\chi_{\l'}(\t',\z'),
\]
where the modular parameters $(\t,\z)$ and $(\t',\z')$ are related by
\beq\label{new-param}
S:
\left|
\begin{array}{l}
\t=-\frac{1}{\t'} \\
\z=-\frac{\z'}{\t'} \end{array} \right.
\quad
{\Longleftrightarrow} \quad
\left|
\begin{array}{l}
\t'=-\frac{1 }{\t} =i \frac{T}{\pi T_0} \\
\z'=\frac{\z}{\t} = \phi \end{array}\right. 
\eeq
and $S_{\l\l'}$ is called the modular $S$-matrix\cite{CFT-book}.
In other words, the $S$-transformation could be used to relate 
the high-temperature and low-temperature limits  of the partition function. 
Note that according to Eq.~(\ref{new-param})  the usual common phase 
factor \cite{cz} in front of the $S$ matrix $S_{\l\l'}$
is trivial since $\exp\left(i\pi\nu_H \Re({\z'}^2/{\t'})\right)=1$.
Now the new modular parameter $q'$ vanishes in the high-temperature limit
\beq\label{q'}
q'=\e^{2\pi i \t'}= \exp\left(-\frac{2T}{T_0} \right) \to 0 \quad
\mathrm{when} \quad T\to \infty
\eeq
so that it would be enough to keep only the leading terms.
 The partition function (\ref{Z_chi}) transforms under the 
$S$-transformation~(\ref{new-param}) as follows
\[
Z(\t,\z)= \sum_{\l=1}^N \chi_{\l}(\t',\z') \sum_{\l'=1}^N S_{\l\l'}
\]
 The quantities $F_\l= \sum_{\l'=1}^N S_{\l\l'}$ shall  play an
 important role in what follows. Since the complete CFT contains a $u(1)$
 factor it turns out that $F_\l \neq 0$  only if the electric charge of 
the representation with label $\l$ is zero, i.e., $Q_{\mathrm{el}}(\l)=0$.
 In particular,
 $F_0=S_{00}\sum_{i=1}^N D_i$, where $D_i=S_{0i}/S_{00}>0$ are the quantum
 dimensions \cite{cgt2000} of the topological excitations of the FQH fluid.

The explicit form of the modular $S$ matrix for all parafermion FQH states 
 could be obtained \cite{NPB-PF_k}
when taking into account  the $\Z_k$ pairing rule and the 
tensor product structure, Eq.~(\ref{struct}), 
of the parafermion coset CFT. Here we only give the final results that 
we need for the computation of the persistent currents in the $k=2,3,4$ 
$\PF_k$ states at higher temperatures.
\subsection{$k=2$}
\label{sec:k2}
The $S$-matrix for the $\PF_2$ model is labeled by the pairs
$(l,\rho)$ where $-1\leq l \leq 2$ and $\rho=0,1$ with the restriction
$\rho \geq l-\rho \mod 2$. Using the explicit form of the $S$ matrix
computed according to the general scheme described in \cite{NPB-PF_k}
we find that
\[
F^{l,\rho}=\sum_{l'=-1}^2 \ \sum_{\rho'\geq l'-\rho'} S^{(l,\rho)}_{(l',\rho')}=
\left\{ \begin{array}{lcl}
	\sqrt{2}+1 & \mathrm{for} & l=0,\ \rho=0 \\
	\sqrt{2}-1 & \mathrm{for} & l=0,\ \rho=1 \\
	0 & & \mathrm{otherwise}
	\end{array}\right.
\]
Notice that the $S$-matrix for  the $\PF_2$ state, which  is known as 
the Pfaffian FQH state, has been explicitly given in a number of papers, see
e.g. Ref.~\cite{cgt} and the references therein.
Thus, the chiral partition function (\ref{Z_PF}) for $k=2$ after the 
$S$ transformation~(\ref{new-param})  can be written as
\beqa
&Z_2(T,\phi) = (\sqrt{2}+1) \chi_{0,0}(\t',\z') +
(\sqrt{2}-1) \chi_{0,1}(\t',\z') = &\nn
&\simeq  \left(\sqrt{2}+1 \right)K_0(\t',2\z';8) +
 \left(\sqrt{2}-1 \right) K_4 (\t',2\z';8) \nonumber , &
\eeqa
where we have dropped out a multiplicative $\z'$-independent term 
proportional to the central
charge, substituted the neutral characters with their leading terms 
$\ch(\L_0+\L_0)(\t')\simeq {q'}^0$,
$\ch(\L_1+\L_1)(\t')\simeq {q'}^{1/2}$ and  ignored  ${q'}^{1/2}$ as compared
to   ${q'}^0$ since $q'\to 0$ when $T\to \infty$. Next, again dropping the
$\eta$ function (\ref{Dede}) and taking only the leading terms in the $K$ 
functions, i.e.,
$n=0$ for $K_0(8)$ and $n=0,-1$ for  $K_4(8)$  one gets
\[
Z_2(T,\phi)  \ \mathop{\simeq}_{T\gg T_0}  \
\left(\sqrt{2}+1 \right)
\left(1+ 2\left( \frac{\sqrt{2}-1}{\sqrt{2}+1}\right) q' \cos(2\pi\z') \right).
\]
 Substituting $q'$ and $\z'$ with  $T$ and $\phi$ according to 
Eq.~(\ref{new-param}) and  Eq.~(\ref{q'}), using the approximation 
$\ln(1+X)\simeq X$ for $X\ll 1$ and taking the derivative with respect to 
$\phi$  according to Eq.~(\ref{pers}) we get
 the following high-temperature asymptotic expression for the persistent
 current ($2\pi^2 k_B T_0/\phi_0= ev_F/L$) in the $\PF_2$ state
 \beqa\label{I_2}
  & I_2(T,\phi)   \ \mathop{\simeq}\limits_{T\gg T_0} \ 
   - \bar{I}_2(T) \sin(2\pi \phi), &   \nn
   & \bar{I}_2(T)  =  \frac{ev_F}{L} \frac{2}{\pi}
		\left(\frac{\sqrt{2}-1}{\sqrt{2}+1}\right)
   \frac{T}{T_0} \exp\left(-2 \frac{T}{T_0}\right). & \quad
 \eeqa
Eq.~(\ref{I_2}) is in agreement with Eqs.~(\ref{high-T}) and (\ref{alpha}) 
with $\alpha_2=2$ being exactly the inverse of the filling factor.
This is an accidental exception from the statement in  Remark 3 
 (and in fact the only one in the entire 
$\PF_k$ hierarchy)
 since for $k=2$ the parafermion CFT dimension $\D^\PF(\L_0+\L_2)=0$.
\subsection{$k=3$}
\label{sec:k3}
For this case the sums of the $S$-matrix elements are found to be 
\cite{cgt2000,NPB-PF_k}
\[
F^{l,\rho}=
\left\{ \begin{array}{rcl}
  \frac{2}{\sqrt{5}} \left(\sin\left(\frac{\pi}{5}\right) +
3 \sin\left(\frac{2\pi}{5}\right) \right)  & \mathrm{for} & l=0,\ \rho=0 \\
  \frac{2}{\sqrt{5}} \left(3\sin\left(\frac{\pi}{5}\right) -
 \sin\left(\frac{2\pi}{5}\right) \right)  & \mathrm{for} & l=0,\ \rho=2 \\
	0 & & \mathrm{otherwise}
	\end{array}\right.
\]
(where $-2\leq l'\leq 2$, $0\leq \rho'\leq 2$ with $\rho'\geq l'-\rho' \mod 3$)
so that the partition function (\ref{Z_PF}) for $k=3$ becomes
\beqa
 Z_3(T,\phi) &=& F^{0,0} \chi_{0,0}(\t',\z') +
F^{0,2} \chi_{0,2}(\t',\z') \simeq \nn
&\simeq& K_0(\t',3\z';15) \left( F^{0,0}{q'}^0+F^{0,2}{q'}^{2/5} \right) + \nn
 &+&  K_5(\t',3\z';15) \left( F^{0,0}{q'}^{2/3}+F^{0,2}{q'}^{1/15} \right) +  \nn
 &+&  K_{-5}(\t',3\z';15) \left( F^{0,0}{q'}^{2/3}+F^{0,2}{q'}^{1/15} \right),
\nonumber
\eeqa
where we have used (skipping the central charge
term in the characters (\ref{ch}) like in the previous subsection) that
$\ch(\L_0+\L_0)(\t')\simeq {q'}^0$,
$\ch(\L_0+\L_1)(\t')=\ch(\L_0+\L_2)(\t')\simeq {q'}^{1/15}$,
$\ch(\L_1+\L_1)(\t')=\ch(\L_2+\L_2)(\t')\simeq {q'}^{2/3}$  and
$\ch(\L_1+\L_2)(\t')\simeq {q'}^{2/5}$. Keeping only the lowest powers in
$q'\to 0$, ignoring the Dedekind function (\ref{Dede}) and taking only $n=0$ in the
$K$ functions  we get
\[
Z_3(T,\phi) \ \mathop{\simeq}_{T\gg T_0}  \
F^{0,0}\left(1  + 2\frac{F^{0,2}}{F^{0,0}} \,
{q'}^{\frac{1}{2} \left[5/3 +2/15 \right]} \cos(2\pi\z')\right)
\]
To get the persistent current in the high-temperature limit we apply the
same scheme as for $k=2$ in Sect.~\ref{sec:k2} which yields
\beqa\label{I_3}
  & I_3(T,\phi)    \mathop{\simeq}\limits_{T\gg T_0} 
   - \bar{I}_3(T) \sin(2\pi \phi), &  \nn
   &  \bar{I}_3(T)  =  \frac{ev_F}{L} \frac{2}{\pi}
    \left(\frac{D-1}{D+1}\right)
   \frac{T}{T_0} \exp\left(- \frac{T}{T_0}\left[ \frac{5}{3} +
\frac{2}{15}\right]\right), & \quad
 \eeqa
where we have transformed the ratio $F^{0,2}/F^{0,0}$ in the form
\[
\frac{F^{0,2}}{F^{0,0}}= \frac{D-1}{D+1}, \quad \mathrm{with}
\quad D=2\cos\left(\frac{\pi}{5}\right)
\]
being the quantum dimension \cite{cgt2000} of the elementary quasihole in 
the $\PF_3$ FQH state.
\subsection{$k=4$}
\label{sec:k4}
The $S$-matrix elements for the $\PF_4$ FQH state have been computed in 
Ref.~\cite{NPB-PF_k}. The quantities $F^{l,\rho}$ can be written as follows 
\[
F^{l,\rho}=\sum_{l'=-2}^3 \ \sum_{\rho'\geq l'-\rho'} 
S^{(l,\rho)}_{(l',\rho')}=
\left\{ \begin{array}{rcl}
  2+\sqrt{3}  & \ \mathrm{for} \ & l=0,\ \rho=0 \\
  2-\sqrt{3} & \ \mathrm{for} \ & l=0,\ \rho=2 \\
   1 & \ \mathrm{for} \ & l=0,\ \rho=3 \\
  0 & & \ \mathrm{otherwise}
	\end{array}\right.
\]
so that the partition function (\ref{Z_PF}) for $k=4$ becomes
\beqa
 && Z_4(T,\phi) =  \nn 
&& = F^{0,0} \chi_{0,0}(\t',\z') +
F^{0,2} \chi_{0,2}(\t',\z') + F^{0,3} \chi_{0,3}(\t',\z')  \nn
&&\simeq K_0(24) \left(F^{0,0}{q'}^{\D^{\PF}(0,0)}+
F^{0,2}{q'}^{\D^{\PF}(2,2)} + F^{0,3}{q'}^{\D^{\PF}(1,3)}\right) + \nn
 &&+  K_6(24) \left(F^{0,0}{q'}^{\D^{\PF}(1,1)}+
 F^{0,2}{q'}^{\D^{\PF}(3,3)} + F^{0,3}{q'}^{\D^{\PF}(0,2)}\right) + \nn
 &&+  K_{-6}(24) \left(F^{0,0}{q'}^{\D^{\PF}(3,3)}+
 F^{0,2}{q'}^{\D^{\PF}(1,1)} + F^{0,3}{q'}^{\D^{\PF}(0,2)}\right),
\nonumber
\eeqa
where $K_l(24)=: K_l(\t',4\z';24)$ and we have used
$\ch(\L_\mu+\L_\rho)(\t')\simeq {q'}^{\D^{\PF}(\mu,\rho)}$,
with
$\D^{\PF}(0,0)= 0 $, $\D^{\PF}(0,2)= 1/12 $,  
$\D^{\PF}(1,1)=\D^{\PF}(3,3)=3/4$,
$\D^{\PF}(1,3)=1/3$,  $\D^{\PF}(2,2)=1$,
again skipping the central charge
term in the characters (\ref{ch}). 
The explicit form of the CFT dimensions $\D^{\PF}(\mu,\rho)$ for the 
parafermion coset $\PF_k$, defined in Eq.~(\ref{struct}), can be found 
in Ref.~\cite{cgt2000}. 

Keeping only the lowest powers in
$q'\to 0$, ignoring the Dedekind function (\ref{Dede}) and taking only 
$n=0$ in the $K$ functions  we get
\[
Z_4(T,\phi) \ \mathop{\simeq}_{T\gg T_0}  \
F^{0,0}   + 2 F^{0,3}{q'}^{\frac{3}{4}+\D^{\PF}(0,2)}  \cos(2\pi\z')
\]
To get the persistent current in the high-temperature limit we apply the
same scheme as in Sect.~\ref{sec:k2} obtaining
\beqa\label{I_4}
  & I_4(T,\phi)    \ \mathop{\simeq}\limits_{T\gg T_0} \ 
   - \bar{I}_4(T) \sin(2\pi \phi), &  \nn
   &  \bar{I}_4(T)  =  \frac{ev_F}{L} \frac{2}{\pi}
    \left(\frac{1}{2+\sqrt{3}}\right)
   \frac{T}{T_0} \ \exp\left(- \frac{T}{T_0}\left[ \frac{3}{2} +
\frac{2}{12}\right]\right). &
\eeqa
Finally, we can extrapolate the high-temperature results 
Eqs.~(\ref{I_2}), (\ref{I_3}) and (\ref{I_4})
for the $k=2$, $3$ and $4$ $\PF_k$ states to the entire parafermion hierarchy.
The universal exponent (\ref{alpha}) for any FQH state can 
be extracted from the CFT operator content and the modular $S$-matrix.
In our case, the analysis in Sects.~\ref{sec:k2}, \ref{sec:k3} and 
\ref{sec:k4}  
implies that  the neutral contribution to the 
exponent is exactly equal to  twice the  minimal CFT dimension of the  
neutral field, that we shall label by $\L_{*}$, whose character is coupled to 
$K_{\pm k+2}(\t',\k\z';k(k+2))$, i.e., 
$\alpha=1/\nu_H+2\D^{\PF}(\L_{*})$, where $\L_{*}$ is 
determined \cite{NPB-PF_k} from the pairing rule condition
\[  
P[\L_{*}]\equiv k+2 \mod k \quad \mathrm{with} \quad 
\D^{\PF}(\L_{*})=\min .
\]
For the $\PF_k$ states $P$ is the $\Z_k$ charge and  
it is easy to see that  $\L_{*}=\L_0+\L_2$ 
for all $k$ in the above equation and that the  formula (\ref{alpha}) 
for $\alpha_k$ is valid for all parafermion FQH states.

Instead of plotting together for comparison  
the analytic high-temperature asymptotic curves and the 
exact numerical ones (the latter being shown on Fig.~\ref{fig:log-decay}), 
we have computed numerically the values of 
$\alpha_k$ and  $I^0_k$  for the $k=2,3,4$ PF states by Least-Squares 
Linear Regression of the logarithmic amplitude 
$\ln \left(\bar{I}_{k}(T)\right)-\ln(T/T_0)$ over 80 points
 in  the interval $6\leq T/T_0 \leq 14$.
It is worth-noting that after subtracting the $\ln(T/T_0)$ contribution from 
 the (numerical) logarithmic amplitude the latter becomes an almost 
perfect straight line as a function of $T$.
 The exact and numerical values
 of the  exponents $\alpha_k$ (the slopes in the logarithmic plots) 
and the asymptotic  amplitudes  $I^0_k$ (the  extrapolated $y$-intercepts) 
for the first several $\PF_k$ states are summarized in  Table~\ref{tab:fit}.
\begin{table}[htb]
	\centering
\caption{The slopes $\alpha_k$ and the $y$-intercepts $I^0_k$ 
(in units $e v_F/L$) of the logarithmic plots in the high-temperature region
for the $k=2,3,4$ PF states. The numerical values have been extracted 
from the numerical data for $6\leq T/T_0 \leq 14$
using the Least-Square linear regression while the  corresponding exact 
quantities have been taken from Eqs.~(\ref{I_2}), (\ref{I_3}) and 
(\ref{I_4}).	\label{tab:fit}}
\begin{tabular}{|c||c|c||c|c|} \hline
  $k$ & \multicolumn{2}{c||}{$\alpha_k$} &
	\multicolumn{2}{c|}{$I^0_k$  $\ [e v_F/L]$} \\
	\cline{2-5}
	& numeric& exact  & numeric & exact \\
	\hline\hline
	$2$ & $2.00071$ & $2$ &  $0.11025$ &
	$(2/\pi)(\sqrt{2}-1)/(\sqrt{2}+1)$ \\
  $3$ & $1.80001$ & $9/5$ & $0.15031$ & 
$(2/\pi)(2\cos(\pi/5)-1)/(2\cos(\pi/5)+1)$\\
	$4$ & $1.66631$ & $5/3$ & $0.16983$ & $(2/\pi)/(2+\sqrt{3})$\\
  \hline
\end{tabular}
\end{table}

\noindent
The amazing overlap between the exact and numerical values of $\alpha_k$ and
$I^0_k$ shown in Table~\ref{tab:fit} gives us the conviction that the 
leading order approximations made in Sects.~\ref{sec:k2}, \ref{sec:k3} 
and \ref{sec:k4} are fairly reasonable and the belief that the 
modular $S$-matrix could  play an important role in the computation 
of certain thermodynamic quantities.

Eq.~(\ref{high-T}) suggests the existence of  another mechanism 
 which reduces the persistent currents amplitudes  at high temperature,
as already mentioned at the end of Sect.~\ref{sec:pers}.
The point is that the finite temperature introduces a length 
scale\cite{geller-loss},
the thermal length $L_T=\hbar v_F/k_B T$ and when, for higher temperatures
this length becomes smaller than the circumference  $L$ of the ring 
the persistent currents vanish exponentially with $L/L_T$ due to 
decoherence effects. 
Following the convention of Ref.~\cite{geller-loss} we take as a 
characteristic  temperature $T_0$ according to Eq.~(\ref{mod_param}) 
so that 
\[
\frac{T}{ T_0}= \pi\frac{L}{L_T}
\]
and the relation with the coherence length $L_{\varphi}$ 
can be  estimated as $L_T\simeq 2\pi L_{\varphi}$ for the 
experimental setup of Ref.~\cite{pers-exp} at $T=15$~mK.
\section{The periods of the persistent currents for the PF states}
\label{sec:period}
The low- and high- temperature asymptotics of the persistent currents
investigated in  Sect.~\ref{sec:low-T} and  Sect.~\ref{sec:high-T}
as well as the numerical calculations performed 
for temperatures in the range  $0.03\leq T/T_0 \leq 14$
unambiguously show that these currents are periodic functions of the AB flux
with period exactly one flux quantum. 
This is in agreement with the Bloch--Byers--Yang theorem
discussed in the Introduction.
On the other hand, our result is at variance with recent theoretical
proposals \cite{ino:comm,kiryu} that the currents have fractional 
period $1/k$ for
the $\PF_k$ states (without contradicting the Bloch theorem).
Note that a fractional periodicity, such as $1/2$, $1/3$ etc., of the 
persistent  currents would be a signal for some broken continuous 
symmetry \cite{deo2}.
However, in incompressible (i.e., effectively $1+1$ dimensional) electron 
systems, described by unitary effective field theories in $1+1$ dimensions, 
a  spontaneous breaking of rotational symmetry  can occur only at
zero temperature \cite{mermin-wagner,coleman,geller:encyclop}. 
One important consequence of our main result confirming that the period 
of the persistent currents in the $\PF_k$ states is 
exactly $1$ is that it  rules out any possibility for a spontaneous breaking
of continuous symmetries at finite temperature.
Nevertheless,  it may be 
possible that the FQH systems undergo phase transitions\cite{geller:encyclop}
 at $T=0$ to some BCS-like phases where the periods could be less than $1$.

\begin{figure}[htb]
\centering
\caption{The position of the maximum of the persistent currents
in the $k=2,3$ and $4$ parafermion FQH states  computed numerically \
for temperatures $0.03\leq T/T_0 \leq 3$.
The inset shows a more detailed plot for lower temperatures.
 \label{fig:phi_max}}
\epsfig{file=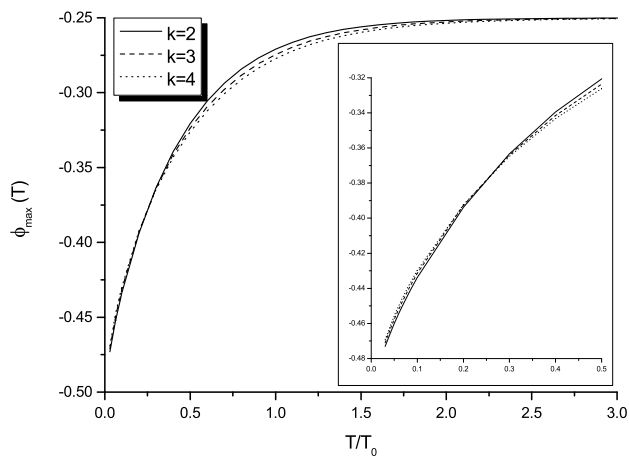,height=7cm}
\end{figure}

For the analysis of the periods of the persistent currents we have 
extracted from the numerical data and plotted
on Fig.~\ref{fig:phi_max} the position of the unique maximum $\phi_{\max}$ 
for $-1/2 \le \phi \le 1/2$
of the persistent currents for the $k=2$, $3$ and $4$ $\PF_k$ states as  
functions of $T$  
in the region $0.03\leq T/T_0 \leq 3$. Obviously, the position of the 
maximum tends  to $-1/2$ for $T\to 0$, in perfect agreement with 
Eq.~(\ref{phi_max})
(we skipped the low-temperature asymptotic curves for $k=2,3,4$ 
that could be plotted from this equation because they become 
indistinguishable from the numerical ones for $T/T_0<0.5$).
For temperatures $T\gg T_0$ the position $\phi_{\max}$ tends quickly 
to $-1/4$, this time  in good agreement with 
Eqs.~(\ref{I_2}), (\ref{I_3}) and (\ref{I_4}).

 Should the period of the currents be smaller than $1$, e.g., $1/k$  for
the $\PF_k$ states, the low-temperature position of the
maximum must be
\[
\left|\phi_{\max}(T) \right| < \frac{1}{2k}, \quad \mathrm{for}\quad
T\ll T_0.
\]
because the persistent currents become for $T\to 0$ odd linear functions of the
flux $\phi$. In order to have period  $1/k$  for the $\PF_k$,
 with $k=2,3,4, \ldots$, it is necessary that   
$\vert\lim_{T\to 0} \phi_{\max}(T)\vert \leq 1/2k$.
This would obviously be  in contradiction with the numerical results
that we have presented here showing that 
\[
\left\vert\lim_{T\to 0} \phi_{\max}(T)\right\vert = \frac{1}{2}.
\]
Therefore, we conclude that our numerical data suggests that 
the periods of the persistent currents for the $\PF_k$ states
cannot be fractional.

Another point we want to stress is that according to 
Eq.~(\ref{phi_max}) the position of the maximum
is independent of the filling factor for $T\ll T_0$ and depends only on 
the ratio between the quasihole's CFT dimension and its electric charge.
\section{Conclusion}
We have proposed a general scheme for the incorporation of the 
Aharonov--Bohm flux in the effective conformal field theory for 
the edge states of a  mesoscopic disk or annulus FQH sample,
which opens the possibility to  study the  mesoscopic effects from 
the CFT point of view.
This approach is based on a special invariance condition (\ref{V})
representing the Laughlin spectral flow and its generalization (\ref{flux}) 
in the CFT framework.

The persistent currents and magnetic susceptibilities 
 can be computed explicitly using the extended
$V$ transformation~(\ref{flux}) and the Cappelli--Zemba  factors~(\ref{CZ}).
We have calculated numerically the persistent currents and magnetic 
susceptibilities for the parafermion FQH states in the second 
Landau level and have analyzed their flux 
periodicities and temperature behavior. The numerical results have been 
presented in several figures. One of our most important observation is
that the periods of the persistent currents and 
magnetic susceptibilities  are exactly one flux quantum for 
all numerically accessible 
temperatures $0.03 \leq T/T_0 \leq 14$, which leaves no room for a 
spontaneous breaking of continuous symmetries.

On the other hand, we have obtained analytic asymptotic expressions,
Eqs.~(\ref{I_max2}), (\ref{kappa_T}) and (\ref{high-T}), for the low-  and 
high-temperature  limits of the partition functions, 
persistent currents amplitudes and zero-field 
magnetic susceptibilities in the parafermion FQH states, which are in 
excellent agreement with our numerical data in the corresponding regimes. 
These expressions  carry important information
about the neutral properties of the FQH fluid and reveal two different 
mechanisms reducing the persistent currents at low- and high- temperatures 
respectively, which are in agreement with the phenomenological 
expectations \cite{pers-exp}.
Our conclusion is that the asymptotic behavior of the persistent currents 
in both regimes in mesoscopic FQH states
 can be used to distinguish between different CFTs for the same 
filling factor. 
In particular, the universal exponent~(\ref{alpha}) and 
the proper quasihole energy~(\ref{proper}) completely characterize the 
neutral properties of the FQH system and 
together with the filling factor
and the minimal quasiparticle electric charge  unambiguously 
determine the FQH universality class.
Surprisingly the universal exponent~(\ref{alpha}) for the 
high-temperature decay of the persistent currents amplitudes 
for the PF states turned out to be 
determined not simply by the filling factor, like in the Laughlin states, 
 but having also a crucial contribution from the neutral sector.

A similar analysis could be performed e.g. for the (principle) Jain  series 
of filling factors where several different CFT have been proposed to 
describe the same electric properties while  differing significantly  
in the neutral sectors.

\begin{acknowledgments}
I would like to thank  Pavel Petkov for giving me access to a
2GHz-class PC as well as Michael Geller and Kazusumi Ino for useful 
discussions.
This work has been partially supported by the FP5-EUCLID Network Program
 of the European Commission under contract HPRN-CT-2002-00325
 and by the Bulgarian National Council for Scientific Research under 
contract F-828.
\end{acknowledgments}

\def\NP{Nucl. Phys. }
\def\PRL{Phys. Rev. Lett.}
\def\PL{Phys. Lett. }
\def\PR{Phys. Rev. }
\def\CMP{Commun. Math. Phys. }
\def\IJMP{Int. J. Mod. Phys. }
\def\JSP{J. Stat. Phys. }
\def\JP{J. Phys. }
\bibliography{Z_k,my}

\begin{thebibliography}{43}
\expandafter\ifx\csname natexlab\endcsname\relax\def\natexlab#1{#1}\fi
\expandafter\ifx\csname bibnamefont\endcsname\relax
  \def\bibnamefont#1{#1}\fi
\expandafter\ifx\csname bibfnamefont\endcsname\relax
  \def\bibfnamefont#1{#1}\fi
\expandafter\ifx\csname citenamefont\endcsname\relax
  \def\citenamefont#1{#1}\fi
\expandafter\ifx\csname url\endcsname\relax
  \def\url#1{\texttt{#1}}\fi
\expandafter\ifx\csname urlprefix\endcsname\relax\def\urlprefix{URL }\fi
\providecommand{\bibinfo}[2]{#2}
\providecommand{\eprint}[2][]{\url{#2}}

\bibitem[{\citenamefont{Thouless}(1998)}]{thouless:top}
\bibinfo{author}{\bibfnamefont{D.~J.} \bibnamefont{Thouless}},
  \emph{\bibinfo{title}{{Toplogical Quantum Numbers in Nonrelativistic
  Physics}}} (\bibinfo{publisher}{World Scientific, Singapore},
  \bibinfo{year}{1998}).

\bibitem[{\citenamefont{Mailly et~al.}(1993)\citenamefont{Mailly, Chapelier,
  and Benoit}}]{pers-exp}
\bibinfo{author}{\bibfnamefont{D.}~\bibnamefont{Mailly}},
  \bibinfo{author}{\bibfnamefont{C.}~\bibnamefont{Chapelier}},
  \bibnamefont{and} \bibinfo{author}{\bibfnamefont{A.}~\bibnamefont{Benoit}},
  \bibinfo{journal}{Phys. Rev. Lett.} \textbf{\bibinfo{volume}{70}},
  \bibinfo{pages}{2020} (\bibinfo{year}{1993}).

\bibitem[{\citenamefont{Geller}(2001)}]{geller:encyclop}
\bibinfo{author}{\bibfnamefont{M.~R.} \bibnamefont{Geller}},
  \bibinfo{journal}{EOLSS Encyclopedia Article}  (\bibinfo{year}{2001}),
  \eprint{cond-mat/0106256}.

\bibitem[{\citenamefont{Aharonov}(1998)}]{QC}
\bibinfo{author}{\bibfnamefont{D.}~\bibnamefont{Aharonov}}, in
  \emph{\bibinfo{booktitle}{Annual Reviews of Computational Physics, vol VI}},
  edited by \bibinfo{editor}{\bibfnamefont{D.}~\bibnamefont{Stauffer}}
  (\bibinfo{publisher}{World Scientific}, \bibinfo{year}{1998}),
  \eprint{quant-ph/9812037}.

\bibitem[{\citenamefont{Deo}(1997)}]{deo}
\bibinfo{author}{\bibfnamefont{P.~S.} \bibnamefont{Deo}},
  \bibinfo{journal}{Phys. Rev.} \textbf{\bibinfo{volume}{B55}},
  \bibinfo{pages}{13795} (\bibinfo{year}{1997}).

\bibitem[{\citenamefont{Altland et~al.}(1996)\citenamefont{Altland, Gefen, and
  Montambaux}}]{altland}
\bibinfo{author}{\bibfnamefont{A.}~\bibnamefont{Altland}},
  \bibinfo{author}{\bibfnamefont{Y.}~\bibnamefont{Gefen}}, \bibnamefont{and}
  \bibinfo{author}{\bibfnamefont{G.}~\bibnamefont{Montambaux}},
  \bibinfo{journal}{Phys. Rev. Lett.} \textbf{\bibinfo{volume}{76}},
  \bibinfo{pages}{1130} (\bibinfo{year}{1996}), \eprint{cond-mat/9511082}.

\bibitem[{\citenamefont{Chakraborty and
  Pietil\"{a}inen}(1994)}]{chakra-pietl-1}
\bibinfo{author}{\bibfnamefont{T.}~\bibnamefont{Chakraborty}} \bibnamefont{and}
  \bibinfo{author}{\bibfnamefont{P.}~\bibnamefont{Pietil\"{a}inen}},
  \bibinfo{journal}{Phys. Rev.} \textbf{\bibinfo{volume}{B50}},
  \bibinfo{pages}{8460} (\bibinfo{year}{1994}).

\bibitem[{\citenamefont{Chakraborty and
  Pietil\"{a}inen}(1995)}]{chakra-pietl-2}
\bibinfo{author}{\bibfnamefont{T.}~\bibnamefont{Chakraborty}} \bibnamefont{and}
  \bibinfo{author}{\bibfnamefont{P.}~\bibnamefont{Pietil\"{a}inen}},
  \bibinfo{journal}{Phys. Rev.} \textbf{\bibinfo{volume}{B52}},
  \bibinfo{pages}{1932} (\bibinfo{year}{1995}).

\bibitem[{\citenamefont{Geller and Loss}(1997)}]{geller-loss}
\bibinfo{author}{\bibfnamefont{M.}~\bibnamefont{Geller}} \bibnamefont{and}
  \bibinfo{author}{\bibfnamefont{D.}~\bibnamefont{Loss}},
  \bibinfo{journal}{Phys. Rev. B} \textbf{\bibinfo{volume}{56}},
  \bibinfo{pages}{9692} (\bibinfo{year}{1997}).

\bibitem[{\citenamefont{Geller et~al.}(1996)\citenamefont{Geller, Loss, and
  Kirczenow}}]{geller-loss-kircz}
\bibinfo{author}{\bibfnamefont{M.}~\bibnamefont{Geller}},
  \bibinfo{author}{\bibfnamefont{D.}~\bibnamefont{Loss}}, \bibnamefont{and}
  \bibinfo{author}{\bibfnamefont{G.}~\bibnamefont{Kirczenow}},
  \bibinfo{journal}{Phys. Rev. Lett.} \textbf{\bibinfo{volume}{77}},
  \bibinfo{pages}{49} (\bibinfo{year}{1996}).

\bibitem[{\citenamefont{Ino}(1998)}]{ino}
\bibinfo{author}{\bibfnamefont{K.}~\bibnamefont{Ino}}, \bibinfo{journal}{Phys.
  Rev. Lett.} \textbf{\bibinfo{volume}{81}}, \bibinfo{pages}{1078}
  (\bibinfo{year}{1998}), \eprint{cond-mat/9803337}.

\bibitem[{\citenamefont{Grayson et~al.}(1998)\citenamefont{Grayson, Tsui,
  Pfeiffer, West, and Chang}}]{grayson}
\bibinfo{author}{\bibfnamefont{M.}~\bibnamefont{Grayson}},
  \bibinfo{author}{\bibfnamefont{D.}~\bibnamefont{Tsui}},
  \bibinfo{author}{\bibfnamefont{L.}~\bibnamefont{Pfeiffer}},
  \bibinfo{author}{\bibfnamefont{K.}~\bibnamefont{West}}, \bibnamefont{and}
  \bibinfo{author}{\bibfnamefont{A.}~\bibnamefont{Chang}},
  \bibinfo{journal}{Phys. Rev. Lett.} \textbf{\bibinfo{volume}{80}},
  \bibinfo{pages}{1062} (\bibinfo{year}{1998}).

\bibitem[{\citenamefont{Fr\"ohlich et~al.}(2001)\citenamefont{Fr\"ohlich,
  Pedrini, Schweigert, and Walcher}}]{fro2000}
\bibinfo{author}{\bibfnamefont{J.}~\bibnamefont{Fr\"ohlich}},
  \bibinfo{author}{\bibfnamefont{B.}~\bibnamefont{Pedrini}},
  \bibinfo{author}{\bibfnamefont{C.}~\bibnamefont{Schweigert}},
  \bibnamefont{and} \bibinfo{author}{\bibfnamefont{J.}~\bibnamefont{Walcher}},
  \bibinfo{journal}{J. Stat. Phys.} \textbf{\bibinfo{volume}{103}},
  \bibinfo{pages}{527} (\bibinfo{year}{2001}), \eprint{cond-mat/0002330}.

\bibitem[{\citenamefont{Fr\"{o}hlich and Kerler}(1991)}]{fro-ker}
\bibinfo{author}{\bibfnamefont{J.}~\bibnamefont{Fr\"{o}hlich}}
  \bibnamefont{and} \bibinfo{author}{\bibfnamefont{T.}~\bibnamefont{Kerler}},
  \bibinfo{journal}{\NP} \textbf{\bibinfo{volume}{B354}}, \bibinfo{pages}{369}
  (\bibinfo{year}{1991}).

\bibitem[{\citenamefont{Fr\"{o}hlich et~al.}(1997)\citenamefont{Fr\"{o}hlich,
  Studer, and Thiran}}]{fro-stu-thi}
\bibinfo{author}{\bibfnamefont{J.}~\bibnamefont{Fr\"{o}hlich}},
  \bibinfo{author}{\bibfnamefont{U.~M.} \bibnamefont{Studer}},
  \bibnamefont{and} \bibinfo{author}{\bibfnamefont{E.}~\bibnamefont{Thiran}},
  \bibinfo{journal}{J. Stat. Phys.} \textbf{\bibinfo{volume}{86}},
  \bibinfo{pages}{821} (\bibinfo{year}{1997}), \eprint{cond-mat/9503113}.

\bibitem[{\citenamefont{Cappelli and Zemba}(1997)}]{cz}
\bibinfo{author}{\bibfnamefont{A.}~\bibnamefont{Cappelli}} \bibnamefont{and}
  \bibinfo{author}{\bibfnamefont{G.~R.} \bibnamefont{Zemba}},
  \bibinfo{journal}{\NP} \textbf{\bibinfo{volume}{B490}}, \bibinfo{pages}{595}
  (\bibinfo{year}{1997}), \eprint{hep-th/9605127}.

\bibitem[{\citenamefont{Laughlin}(1983)}]{laugh}
\bibinfo{author}{\bibfnamefont{R.}~\bibnamefont{Laughlin}},
  \bibinfo{journal}{Phys. Rev. Lett.} \textbf{\bibinfo{volume}{50}},
  \bibinfo{pages}{1395} (\bibinfo{year}{1983}).

\bibitem[{\citenamefont{Geller and Vignale}(1995)}]{geller-vignale}
\bibinfo{author}{\bibfnamefont{M.~R.} \bibnamefont{Geller}} \bibnamefont{and}
  \bibinfo{author}{\bibfnamefont{G.}~\bibnamefont{Vignale}},
  \bibinfo{journal}{Phys. Rev. B} \textbf{\bibinfo{volume}{52}},
  \bibinfo{pages}{14137} (\bibinfo{year}{1995}), \eprint{cond-mat/9412028}.

\bibitem[{\citenamefont{Geller}(2002)}]{michael}
\bibinfo{author}{\bibfnamefont{M.}~\bibnamefont{Geller}},
  \bibinfo{journal}{private communication}  (\bibinfo{year}{2002}).

\bibitem[{\citenamefont{Read and Rezayi}(1998)}]{rr}
\bibinfo{author}{\bibfnamefont{N.}~\bibnamefont{Read}} \bibnamefont{and}
  \bibinfo{author}{\bibfnamefont{E.}~\bibnamefont{Rezayi}},
  \bibinfo{journal}{\PR} \textbf{\bibinfo{volume}{B59}}, \bibinfo{pages}{8084}
  (\bibinfo{year}{1998}).

\bibitem[{\citenamefont{Pan et~al.}(1999)\citenamefont{Pan, Xia, Shvarts,
  Adams, St\"ormer, Tsui, Pfeiffer, Baldwin, and West}}]{pan}
\bibinfo{author}{\bibfnamefont{W.}~\bibnamefont{Pan}},
  \bibinfo{author}{\bibfnamefont{J.-S.} \bibnamefont{Xia}},
  \bibinfo{author}{\bibfnamefont{V.}~\bibnamefont{Shvarts}},
  \bibinfo{author}{\bibfnamefont{D.~E.} \bibnamefont{Adams}},
  \bibinfo{author}{\bibfnamefont{H.~L.} \bibnamefont{St\"ormer}},
  \bibinfo{author}{\bibfnamefont{D.~C.} \bibnamefont{Tsui}},
  \bibinfo{author}{\bibfnamefont{L.~N.} \bibnamefont{Pfeiffer}},
  \bibinfo{author}{\bibfnamefont{K.~W.} \bibnamefont{Baldwin}},
  \bibnamefont{and} \bibinfo{author}{\bibfnamefont{K.~W.} \bibnamefont{West}},
  \bibinfo{journal}{Phys. Rev. Lett.} \textbf{\bibinfo{volume}{83}},
  \bibinfo{pages}{3530} (\bibinfo{year}{1999}), \eprint{cond-mat/9907356}.

\bibitem[{\citenamefont{Cappelli et~al.}(2001)\citenamefont{Cappelli, Georgiev,
  and Todorov}}]{cgt2000}
\bibinfo{author}{\bibfnamefont{A.}~\bibnamefont{Cappelli}},
  \bibinfo{author}{\bibfnamefont{L.}~\bibnamefont{Georgiev}}, \bibnamefont{and}
  \bibinfo{author}{\bibfnamefont{I.}~\bibnamefont{Todorov}},
  \bibinfo{journal}{Nucl. Phys.} \textbf{\bibinfo{volume}{B599}},
  \bibinfo{pages}{499} (\bibinfo{year}{2001}), \eprint{hep-th/0009229}.

\bibitem[{\citenamefont{Mermin and Wagner}(1966)}]{mermin-wagner}
\bibinfo{author}{\bibfnamefont{N.}~\bibnamefont{Mermin}} \bibnamefont{and}
  \bibinfo{author}{\bibfnamefont{H.}~\bibnamefont{Wagner}},
  \bibinfo{journal}{Phys. Rev. Lett.} \textbf{\bibinfo{volume}{17}},
  \bibinfo{pages}{1133} (\bibinfo{year}{1966}).

\bibitem[{\citenamefont{Coleman}(1973)}]{coleman}
\bibinfo{author}{\bibfnamefont{S.}~\bibnamefont{Coleman}},
  \bibinfo{journal}{Commun. Math. Phys.} \textbf{\bibinfo{volume}{31}},
  \bibinfo{pages}{259} (\bibinfo{year}{1973}).

\bibitem[{\citenamefont{Georgiev}(2003{\natexlab{a}})}]{5-2}
\bibinfo{author}{\bibfnamefont{L.}~\bibnamefont{Georgiev}},
  \bibinfo{journal}{Nucl. Phys.} \textbf{\bibinfo{volume}{B651}},
  \bibinfo{pages}{331} (\bibinfo{year}{2003}{\natexlab{a}}),
  \eprint{hep-th/0108173}.

\bibitem[{\citenamefont{Georgiev}(2003{\natexlab{b}})}]{NPB-PF_k}
\bibinfo{author}{\bibfnamefont{L.}~\bibnamefont{Georgiev}},
  \bibinfo{journal}{work in progress}  (\bibinfo{year}{2003}{\natexlab{b}}).

\bibitem[{\citenamefont{Wen}(1995)}]{wen-top}
\bibinfo{author}{\bibfnamefont{X.-G.} \bibnamefont{Wen}},
  \bibinfo{journal}{Adv. Phys.} \textbf{\bibinfo{volume}{44}},
  \bibinfo{pages}{405} (\bibinfo{year}{1995}).

\bibitem[{\citenamefont{Georgiev}(2002)}]{gaps}
\bibinfo{author}{\bibfnamefont{L.}~\bibnamefont{Georgiev}},
  \bibinfo{journal}{Nucl. Phys.} \textbf{\bibinfo{volume}{B626}},
  \bibinfo{pages}{415} (\bibinfo{year}{2002}), \eprint{cond-mat/0102451}.

\bibitem[{\citenamefont{\uppercase{D}i Francesco
  et~al.}(1997)\citenamefont{\uppercase{D}i Francesco, Mathieu, and
  S\'en\'echal}}]{CFT-book}
\bibinfo{author}{\bibfnamefont{P.}~\bibnamefont{\uppercase{D}i Francesco}},
  \bibinfo{author}{\bibfnamefont{P.}~\bibnamefont{Mathieu}}, \bibnamefont{and}
  \bibinfo{author}{\bibfnamefont{D.}~\bibnamefont{S\'en\'echal}},
  \emph{\bibinfo{title}{Conformal Field Theory}}
  (\bibinfo{publisher}{Springer--Verlag, New York}, \bibinfo{year}{1997}).

\bibitem[{\citenamefont{Cappelli et~al.}(1996)\citenamefont{Cappelli,
  Trugenberger, and Zemba}}]{ctz3}
\bibinfo{author}{\bibfnamefont{A.}~\bibnamefont{Cappelli}},
  \bibinfo{author}{\bibfnamefont{C.}~\bibnamefont{Trugenberger}},
  \bibnamefont{and} \bibinfo{author}{\bibfnamefont{G.}~\bibnamefont{Zemba}},
  \bibinfo{journal}{Annals Phys.} \textbf{\bibinfo{volume}{246}},
  \bibinfo{pages}{86} (\bibinfo{year}{1996}), \eprint{cond-mat/9407095}.

\bibitem[{\citenamefont{Georgiev and Todorov}(1998)}]{gt}
\bibinfo{author}{\bibfnamefont{L.}~\bibnamefont{Georgiev}} \bibnamefont{and}
  \bibinfo{author}{\bibfnamefont{I.}~\bibnamefont{Todorov}},
  \bibinfo{journal}{J. Math. Phys.} \textbf{\bibinfo{volume}{39}},
  \bibinfo{pages}{5762} (\bibinfo{year}{1998}), \eprint{hep-th/9710134}.

\bibitem[{\citenamefont{Ino}(2000)}]{ino2}
\bibinfo{author}{\bibfnamefont{K.}~\bibnamefont{Ino}}, \bibinfo{journal}{Phys.
  Rev.} \textbf{\bibinfo{volume}{B62}}, \bibinfo{pages}{6936}
  (\bibinfo{year}{2000}), \eprint{cond-mat/0008094}.

\bibitem[{\citenamefont{Kiryu}(2002)}]{kiryu}
\bibinfo{author}{\bibfnamefont{H.}~\bibnamefont{Kiryu}},
  \bibinfo{journal}{Phys. Rev.} \textbf{\bibinfo{volume}{B65}},
  \bibinfo{pages}{113407} (\bibinfo{year}{2002}).

\bibitem[{\citenamefont{Byers and Yang}(1961)}]{byers-yang}
\bibinfo{author}{\bibfnamefont{N.}~\bibnamefont{Byers}} \bibnamefont{and}
  \bibinfo{author}{\bibfnamefont{C.}~\bibnamefont{Yang}},
  \bibinfo{journal}{Phys. Rev. Lett.} \textbf{\bibinfo{volume}{7}},
  \bibinfo{pages}{46} (\bibinfo{year}{1961}).

\bibitem[{\citenamefont{Morf and Halperin}(1986)}]{morf:proper}
\bibinfo{author}{\bibfnamefont{R.}~\bibnamefont{Morf}} \bibnamefont{and}
  \bibinfo{author}{\bibfnamefont{B.}~\bibnamefont{Halperin}},
  \bibinfo{journal}{Phys. Rev. B} \textbf{\bibinfo{volume}{33}},
  \bibinfo{pages}{2221} (\bibinfo{year}{1986}).

\bibitem[{\citenamefont{Deblock et~al.}(2002)\citenamefont{Deblock, Bel,
  Reulet, Bouchiat, and Mailly}}]{diamag}
\bibinfo{author}{\bibfnamefont{R.}~\bibnamefont{Deblock}},
  \bibinfo{author}{\bibfnamefont{R.}~\bibnamefont{Bel}},
  \bibinfo{author}{\bibfnamefont{B.}~\bibnamefont{Reulet}},
  \bibinfo{author}{\bibfnamefont{H.}~\bibnamefont{Bouchiat}}, \bibnamefont{and}
  \bibinfo{author}{\bibfnamefont{D.}~\bibnamefont{Mailly}},
  \bibinfo{journal}{Phys. Rev. Lett.} \textbf{\bibinfo{volume}{89}},
  \bibinfo{pages}{206803} (\bibinfo{year}{2002}).

\bibitem[{\citenamefont{Loss}(1992)}]{loss-parity}
\bibinfo{author}{\bibfnamefont{D.}~\bibnamefont{Loss}}, \bibinfo{journal}{Phys.
  Rev. Lett.} \textbf{\bibinfo{volume}{69}}, \bibinfo{pages}{343}
  (\bibinfo{year}{1992}).

\bibitem[{\citenamefont{Loss and Goldbart}(1991)}]{loss-goldbart}
\bibinfo{author}{\bibfnamefont{D.}~\bibnamefont{Loss}} \bibnamefont{and}
  \bibinfo{author}{\bibfnamefont{P.}~\bibnamefont{Goldbart}},
  \bibinfo{journal}{Phys. Rev.} \textbf{\bibinfo{volume}{B43}},
  \bibinfo{pages}{13762} (\bibinfo{year}{1991}).

\bibitem[{\citenamefont{Kim et~al.}(2002)\citenamefont{Kim, Cho, Kim, and
  Nahm}}]{kim}
\bibinfo{author}{\bibfnamefont{M.~D.} \bibnamefont{Kim}},
  \bibinfo{author}{\bibfnamefont{S.~Y.} \bibnamefont{Cho}},
  \bibinfo{author}{\bibfnamefont{C.~K.} \bibnamefont{Kim}}, \bibnamefont{and}
  \bibinfo{author}{\bibfnamefont{K.}~\bibnamefont{Nahm}},
  \bibinfo{journal}{Phys. Rev.} \textbf{\bibinfo{volume}{B66}},
  \bibinfo{pages}{193308} (\bibinfo{year}{2002}), \eprint{cond-mat/0111102}.

\bibitem[{\citenamefont{Geller et~al.}(1997)\citenamefont{Geller, Loss, and
  Kirczenow}}]{geller-loss-kircz-1}
\bibinfo{author}{\bibfnamefont{M.}~\bibnamefont{Geller}},
  \bibinfo{author}{\bibfnamefont{D.}~\bibnamefont{Loss}}, \bibnamefont{and}
  \bibinfo{author}{\bibfnamefont{G.}~\bibnamefont{Kirczenow}},
  \bibinfo{journal}{Superlattices Microstruct.} \textbf{\bibinfo{volume}{21}},
  \bibinfo{pages}{5110} (\bibinfo{year}{1997}).

\bibitem[{\citenamefont{Cappelli et~al.}(1999)\citenamefont{Cappelli, Georgiev,
  and Todorov}}]{cgt}
\bibinfo{author}{\bibfnamefont{A.}~\bibnamefont{Cappelli}},
  \bibinfo{author}{\bibfnamefont{L.}~\bibnamefont{Georgiev}}, \bibnamefont{and}
  \bibinfo{author}{\bibfnamefont{I.}~\bibnamefont{Todorov}},
  \bibinfo{journal}{Commun. Math. Phys.} \textbf{\bibinfo{volume}{205}},
  \bibinfo{pages}{657} (\bibinfo{year}{1999}), \eprint{hep-th/9810105}.

\bibitem[{\citenamefont{Ino}(2003)}]{ino:comm}
\bibinfo{author}{\bibfnamefont{K.}~\bibnamefont{Ino}}, \bibinfo{journal}{to be
  published}  (\bibinfo{year}{2003}).

\bibitem[{\citenamefont{Deo et~al.}(2001)\citenamefont{Deo, Koskinen, Koskinen,
  and Manninen}}]{deo2}
\bibinfo{author}{\bibfnamefont{P.~S.} \bibnamefont{Deo}},
  \bibinfo{author}{\bibfnamefont{P.}~\bibnamefont{Koskinen}},
  \bibinfo{author}{\bibfnamefont{M.}~\bibnamefont{Koskinen}}, \bibnamefont{and}
  \bibinfo{author}{\bibfnamefont{M.}~\bibnamefont{Manninen}}
  (\bibinfo{year}{2001}), \eprint{cond-mat/0112224}.

\end{thebibliography}

\end{document}